\documentclass[lettersize,journal]{IEEEtran}
\usepackage{amsmath,amsfonts}
\usepackage{algorithmic}
\usepackage{algorithm}
\usepackage{array}
\usepackage[font=normalsize,labelfont=sf,textfont=sf]{subcaption}
\usepackage{textcomp}
\usepackage{stfloats}
\usepackage{url}
\usepackage{verbatim}
\usepackage{graphicx}
\usepackage{xcolor}
\usepackage{bbm}
\usepackage{amsthm}
\usepackage[noadjust]{cite}
\usepackage{url}

\newfont{\bbb}{msbm10 scaled 700}

\newcommand{\EEE}{\mbox{\bbb E}}

\newfont{\bb}{msbm10 scaled 1100}
\newcommand{\CC}{\mbox{\bb C}}
\newcommand{\PP}{\mbox{\bb P}}
\newcommand{\RR}{\mbox{\bb R}}

\newcommand{\EE}{\mbox{\bb E}}

\newcommand{\HH}{\mbox{\bb H}}

\newcommand{\UU}{\mbox{\bb U}}

\newcommand{\yy}{\mathbbm{y}}
\newcommand{\xx}{\mathbbm{x}}
\newcommand{\zz}{\mathbbm{z}}
\newcommand{\sss}{\mathbbm{s}}

\newcommand{\hh}{\mathbbm{h}}
\newcommand{\uu}{\mathbbm{u}}

\usepackage[mathscr]{euscript}

\newcommand{\av}{{\bf a}}

\newcommand{\ev}{{\bf e}}

\newcommand{\hv}{{\bf h}}

\newcommand{\rv}{{\bf r}}
\newcommand{\sv}{{\bf s}}

\newcommand{\uv}{{\bf u}}
\newcommand{\wv}{{\bf w}}
\newcommand{\vv}{{\bf v}}
\newcommand{\xv}{{\bf x}}
\newcommand{\yv}{{\bf y}}
\newcommand{\zv}{{\bf z}}
\newcommand{\zerov}{{\bf 0}}

\newcommand{\Fm}{{\bf F}}
\newcommand{\Gm}{{\bf G}}

\newcommand{\Id}{{\bf I}}

\newcommand{\Ym}{{\bf Y}}
\newcommand{\Zm}{{\bf Z}}

\newcommand{\Bc}{{\cal B}}
\newcommand{\Cc}{{\cal C}}

\newcommand{\Ec}{{\cal E}}

\newcommand{\Gc}{{\cal G}}

\newcommand{\Kc}{{\cal K}}
\newcommand{\Lc}{{\cal L}}
\newcommand{\Mc}{{\cal M}}
\newcommand{\Nc}{{\cal N}}

\newcommand{\Qc}{{\cal Q}}

\newcommand{\Sc}{{\cal S}}
\newcommand{\Tc}{{\cal T}}
\newcommand{\Uc}{{\cal U}}

\newcommand{\Xc}{{\cal X}}
\newcommand{\Yc}{{\cal Y}}

\newcommand{\nuv}{\hbox{\boldmath$\nu$}}

\newcommand{\zetav}{\hbox{\boldmath$\zeta$}}
\newcommand{\phiv}{\hbox{\boldmath$\phi$}}

\newcommand{\Gammam}{\hbox{\boldmath$\Gamma$}}

\newcommand{\Sigmam}{\hbox{\boldmath$\Sigma$}}

\newcommand{\diag}{{\hbox{diag}}}

\newcommand{\trace}{{\hbox{tr}}}

\newcommand{\eqdef}{\stackrel{\Delta}{=}}

\newcommand{\herm}{{\sf H}}

\newcommand{\transp}{{\sf T}}

\newcommand{\SINR}{{\sf SINR}}
\newcommand{\SNR}{{\sf SNR}}

\newcommand{\taudmrs}{\tau_p}

\newcommand{\fhratekUL}{B_{\ell, k}}
\newcommand{\fhratekDL}{R^{\rm dl}_k}
\usepackage{xparse}

\NewDocumentCommand{\ceil}{s O{} m}{%
	\IfBooleanTF{#1} 
	{\left\lceil#3\right\rceil} 
	{#2\lceil#3#2\rceil} 
}

\newcommand{\defeq}{ \stackrel{\triangle}{=} }

\usepackage[font={small}]{caption}

\newtheorem{example}{Example}
\newtheorem{definition}{Definition}
\newtheorem{remark}{Remark}

\newtheorem{lemma}{Lemma}

\allowdisplaybreaks

\begin{document}

\title{Joint Fronthaul Load Balancing and Computation Resource Allocation in Cell-Free User-Centric 
	Massive MIMO Networks}

\author{Zhiyang Li, Fabian G\"ottsch,~\IEEEmembership{Student Member, IEEE,} Siyao Li, Ming Chen,~\IEEEmembership{Member, IEEE,} Giuseppe Caire,~\IEEEmembership{Fellow, IEEE}
	
	\thanks{Part of this paper was accepted for presentation at the 2024 IEEE International Conference on Communications.  
		}
	\thanks{ Fabian G\"ottsch and Giuseppe Caire are with the Faculty of EECS, Technische Universit{\"a}t Berlin, 10587 Berlin, Germany. Email: {\tt \{fabian.goettsch, caire\}@tu-berlin.de}}
	\thanks{ Siyao Li is with the Department of Electrical Engineering, University of Alaska Anchorage, 99508, Alaska, USA. Email: {\tt sli15@alaska.edu}}
	\thanks{ Zhiyang Li and Ming Chen are with the National Mobile Communications Research Laboratory, Southeast University, Nanjing 211111, China. Email: {\tt 
			\{lizhiyang, chenming\}@seu.edu.cn}}
}

\maketitle

\begin{abstract}
	We consider {\em scalable} cell-free massive multiple-input multiple-output networks under an
	{\em open radio access network} paradigm comprising user equipments (UEs), radio units (RUs), and decentralized processing units (DUs). 
	UEs are served by dynamically allocated user-centric clusters of RUs. The corresponding 
	{\em cluster processors} (implementing the physical layer for each user) 
	are hosted by the DUs as {\em software-defined virtual network functions}. 
	Unlike the current literature, mainly focused on the characterization of the user rates 
	under unrestricted fronthaul communication and computation, 
	in this work we explicitly take into account the fronthaul topology, the limited fronthaul communication capacity, 
	and computation constraints at the DUs. 
	In particular, we systematically address the new problem of joint fronthaul load balancing and 
	allocation of the computation resource. As a consequence of our new optimization framework, we present 
	representative numerical results highlighting the existence of an optimal number of quantization bits in the analog-to-digital conversion at the RUs. 
\end{abstract}


\section{Introduction}   \label{intro}
Wireless data traffic has witnessed a remarkable surge over the past few decades, culminating with the
deployment of fifth-generation (5G) mobile/wireless networks \cite{5Gnetworks}. 
To address the escalating traffic demand, 5G has relied on the densification of network infrastructure \cite{Ultra-Dense} 
coupled with the implementation of massive multiple-input multiple-output (mMIMO) techniques \cite{Noncooperative,ScalingUp,Larsson-book} to enhance the per-cell spectral efficiency (SE). 
For the next standardization cycle (6G), even more substantial demands in terms of traffic volume are envisaged, thereby mandating the development of innovative and efficient solutions.
In this context, cell-free mMIMO (CF-mMIMO) emerges as a highly promising approach to meet the 
needs of 6G networks, capitalizing on the benefits of both infrastructure densification and mMIMO 
(e.g., see \cite{Cell-Free,Precoding2017, bjo2020, DemirFoundation,Subspace-Based} and references therein).
In parallel, the open radio access network (O-RAN) paradigm has gained significant attention by defining open interfaces between radio units (RUs) and decentralized baseband processing units (DUs). 
O-RAN enables multi-vendor interoperability and the implementation of the physical layer (PHY) 
as a software-defined virtual network function running on general-purpose hardware such as CPUs and GPUs, 
 and it is ideally suited (although not strictly necessary) for the implementation of 
CF-mMIMO (see, e.g., \cite{demir2023cellfree, ranjbar2022cell, beerten2022user}).

A fundamental aspect that distinguishes CF-mMIMO from traditional Cloud Radio Access Network (C-RAN) systems  \cite{wu2015cloud}, where distributed antennas are connected to a centralized processing node, is its {\em scalability} \cite{bjo2020,DemirFoundation}. In short, this is the ability to handle a constant density of user equipments (UEs), RUs, and DUs as the network area grows infinitely, while the computation capacity at each network node and the communication capacity in each network link remain constant. 
Since users are served by a finite-size \textit{user-centric cluster} of RUs 
that depend on the user locations and are not a priori determined,
an essential requirement for scalability is that RUs and DUs are connected by a flexible fronthaul network, 
able to route the data traffic between RUs and DUs. In this work, we assume that such flexible fronthaul is formed by capacitated point-to-point wired or wireless links
and intermediate routing-capable nodes, referred to for brevity as {\em routers} (see  Fig.~\ref{fig1}). 

Each DU has a finite computation capacity and can host up to a given maximum number of {\em cluster processors}. 
A cluster processor is a {\em software-defined virtual network function} 
(running on the general-purpose hardware of the DU) that 
implements the PHY for the corresponding UE. This includes pilot-based channel estimation in the uplink (UL), 
computation of the linear receiver vector for mMIMO interference mitigation and UL detection, and UL channel decoding, 
as well as computation of the downlink (DL) precoding vectors and of the precoded signals to be transmitted jointly by the cluster of RUs in the DL. We assume time division duplexing (TDD) and UL/DL channel reciprocity, such that 
the UL channel state information can also be used for the  DL precoder computation \cite{DemirFoundation}. 
In particular, we consider the PHY defined in \cite{Subspace-Based}, which is well-known and widely investigated. 
Notice however that our general optimization framework applies for any suitably defined PHY layer, 
as in the many variants  presented in the CF-mMIMO literature (see \cite{DemirFoundation} and references therein). 

In order to alleviate the fronthaul load,  the key system design aspects are:
1) optimal routing of the UL (from RUs to DUs) and DL (from DUs to RUs) data traffic such that 
the maximum link load is minimized; 2) optimal placement of the cluster processor in the DUs, such that 
each DU does not violate its maximum computation capacity; 3) optimization of the number of 
{\em bits per signal dimension}~\footnote{It is well-known that finite energy bandpass 
	signals of single-sided bandwidth $W$ and approximate duration $T$ span a signal space of dimension
	$\approx WT$ over the complex field $\CC$ for large product $WT$ \cite{gallager1968information}. 
	As usual in communication theory, taking this limit with equality, we identify bit per signal dimensions, i.e., per baseband complex sample, 
	with the spectral efficiency in bit/s/Hz.}
to represent the received signal in the UL (from RUs to DUs) and the 
user information messages in the DL (from DUs to RUs). 

For a given network topology, the load balancing (via routing) and the allocation of computation resources must be jointly optimized. 
In this paper we formulate this joint optimization problem in the form of  mixed-integer linear programs (MILPs), 
which can be efficiently solved even for fairly large networks. We would like to stress the fact that the problem formulation and 
its optimization framework represent a novel contribution that has not yet been addressed in the CF-mMIMO literature.

\subsection{Related works}

In C-RAN  architectures centralized around a single processing unit, extensive research has delved into fronthaul capacity from an information-theoretic perspective, exemplified in works like \cite{zhou2016optimal,lee2016multivariate,patil2019generalized,aguerri2019capacity,liu2016uplink}. 
These studies model the UL and DL as multiaccess-relay and broadcast-relay channels with RUs serving as ``oblivious'' relays, where rigorous achievability region results and UL/DL duality results have been provided under idealized conditions of perfect channel state information (CSI). 
Moreover, CF-mMIMO networks have been investigated also with respect to  
data quantization and forwarding schemes (e.g., see
\cite{Manijeh_2018, Bashar_2018, Limited-Backhaul, masoumi2019performance,bashar2019maxmin}). 
Related optimization problems include the pursuit of max-min user PHY rates within limited fronthaul capacity constraints \cite{boroujerdi2019cell} and quantization bit limitations \cite{hu2019cell}, allowing each RU to employ a distinct number of quantization bits.  Moving forward, \cite{Ettefagh2023} presents finite-length coding achievable rates in both UL and DL for fully digital quantized multiuser MIMO systems, focusing on scenarios involving a single DU, and 
\cite{Femenias_2020} offers analytical closed-form expressions for achievable user PHY rates in both UL and DL of CF-mMIMO networks, employing low-resolution Analog-to-Digital Converters (ADCs).

Other relevant research endeavors include  the exploration of a user-centric approach aimed at minimizing the fronthaul load \cite{Buzzi2017cellfree}, and the investigation of joint power allocation and RU scheduling within UL limited-capacity fronthaul CF-mMIMO networks \cite{Guenach2021powerControl}, considering constraints such as maximum user Tx power and an upper bound on shared fronthaul bandwidth. 
	
	It is worth noting that all the aforementioned studies primarily focus on C-RAN-type networks with a single ``giant'' central processing unit, to which all RUs are connected,  i.e., not considering fronthaul routing. As such, none of the above mentioned works effectively considers scalable CF-mMIMO networks (as defined in  \cite{bjo2020,DemirFoundation}). 
	In contrast, we are the first to notice the essential problem of fronthaul load balancing, which is obviously related to 
where (in which DU)  
the PHY processors (referred to as ``cluster processors'') are placed. 
Hence, for the first time, our paper addresses the problem of the joint load balancing and computation resource allocation in such networks, which is an essential and {\em structural} 
problem intimately connected with the concept of user-centric clusters and 
scalable CF architectures, and nevertheless was not noticed and addressed in the existing literature. 

\subsection{Contributions}

The main contributions of the paper are summarized as follows.

\begin{itemize}
	\item 
	We propose a comprehensive high-level system model that captures the fronthaul data flow in 
	CF-mMIMO networks. 
	\item { We formulate the joint fronthaul load balancing and cluster processor placement problem with the objective of minimizing the maximum link load (bottleneck link). This mixed-integer linear problem can be solved using standard optimization tools \cite{elie2019}. We address two versions of the fundamental problem, depending on whether the fronthaul has full-duplex or half-duplex links, accommodating an arbitrary duty-cycle 
	of the time-division duplex (TDD) UL and DL slots.}
	\item  We use information theoretic rate distortion theory to analytically model the effect of analog-to-digital (A/D) quantization and 
	study the tradeoff between the optimized fronthaul capacity and the achievable rate on the radio access network (UL/DL user rates) as a function of the received signal quantization (in the UL) and of the precoded 
	signal representation (in the DL).    
	\item We illustrate our approach by evaluating through simulation 
		representative network topologies employing the 3GPP urban microcell street canyon pathloss model. 
		We discuss the effect of the A/D quantization and fronthaul compression, and show results 
		in terms of the achievable PHY user rates versus (optimized) fronthaul load.
\end{itemize}

The rest of the paper is organized as follows. Section \ref{sec:sys} introduces the system model including network topology, channel information, and uplink and downlink transmissions. Section \ref{sec:UL-Distributed} formulates  the max link load minimization problem for UL and DL under fronthaul full-duplex and half-duplex assumptions.  Section \ref{sec:simulation} presents illustrative numerical results and discusses the effect of the different system parameters. Section \ref{sec:conclusion}  concludes this work.

\section{System Model} \label{sec:sys}

\subsection{Network topology and MIMO channel statistics}  \label{network}

We consider a cell-free network with $K$ single-antenna UEs, $L$ RUs, $Q$ routers, and $N$ DUs. 
Each RU is equipped with $M$ antennas. 
The sets of UEs, RUs, routers, and DUs are denoted by 
$\mathcal K=\{1,\dots, K\}$, $\mathcal L=\{1,\dots, L\}$, $\mathcal Q=\{1,\dots, Q\}$, and $\mathcal N=\{1,\dots, N\}$, respectively.
Each UE is associated to a user-centric cluster of surrounding RUs, according to 
a cluster formation scheme described later. 
The RUs, routers and DUs are connected via a fronthaul mesh network formed by 
non-interfering error-free digital links. 

The topology of the radio access segment of the network is defined by the UE-RU associations. This is expressed by a bipartite graph $\Gc_{\rm ran}(\Kc, \Lc, \Ec_{\rm ran})$ with two classes of nodes (UEs and RUs).  The cluster of RUs serving user $k \in \Kc$ is denoted by $\Cc_k  \subseteq \Lc$. 
The set of UEs served by RU $\ell \in \Lc$ is denoted by $\Uc_\ell \subseteq \Kc$.  Notice that, for any $\ell \in \Lc$ and $k \in \Kc$,  $\ell \in \Cc_k$ iff $k \in \Uc_\ell$.  The set of edges of $\Ec_{\rm ran}$ is such that $(\ell, k) \in \Ec_{\rm ran}$ iff $\ell \in \Cc_k$ (or, equivalently, iff $k \in \Uc_\ell$).
The fronthaul network is also represented as a graph $\Gc_{\rm front}(\Lc, \Qc, \Nc, \Ec _{\rm front})$ 
with capacitated edges.
Notice that $\Gc_{\rm front}(\Lc, \Qc, \Nc, \Ec _{\rm front})$ is a general (oriented) graph, i.e.,  RUs may communicate with other RUs, directly with DUs, or with routers. The overall network topology is described by the union of
$\Gc_{\rm ran}(\Kc, \Lc, \Ec_{\rm ran})$ and $\Gc_{\rm front}(\Lc, \Qc, \Nc, \Ec _{\rm front})$ 
obtained by merging the common nodes $\Lc$.  

The network operates the UL and DL in TDD, such that the UL and DL channel coefficients remain constant (channel reciprocity) over blocks of $T$ signal dimensions (time-frequency channel uses), referred to as ``resource blocks''.~\footnote{This block-fading model is standard and it is used in countless works on 
	multiuser MIMO, massive MIMO, and cell-free systems (e.g., see \cite{Noncooperative,Larsson-book,DemirFoundation} and references therein). The  model is justified  by considering OFDM modulation, such that the channel coefficients between any UE and RU antenna pair is random but constant over a certain number of OFDM symbols in time $N_{\rm symb}$ 
	and a certain number of OFDM subcarriers $N_{\rm sub}$, where  $T = N_{\rm symb}N_{\rm sub}$ time-frequency channel uses. 
	We borrow the term ``resource-block'' from the 5GNR standard \cite{3gpp38901}.}  
	In general, the UL and DL traffic demand may be significantly imbalanced. This is handled by 
	allocating a different fraction of resource blocks to UL and DL in the TDD frame. 
	We denote by ${ {\gamma_{\rm DL}}} \in (0,1)$ the resource fraction allocated to the DL, and
	as a consequence $(1 - { {\gamma_{\rm DL}}})$ is allocated to the UL. { Note that this parameter is set by the TDD scheduler and depends on the overall traffic demand imbalance between UL and DL.}
	
At any given resource block, we let $\mathbb{H} \in \mathbb{C}^{LM \times K}$ denote the overall channel matrix collecting the channel coefficients between all UEs and all the RU antennas. 
Since channel estimation, UL detection, and DL precoding, are defined on a per-resource-block basis, 
and since the rate expressions depend only on the first-order marginal statistics of $\mathbb{H}$ and not on its correlation over 
multiple resource blocks (as long as it satisfies the usual assumptions of stationarity and ergodicity), for simplicity we
focus on a generic resource block and we omit the resource block index.  
The channel matrix $\mathbb{H}$ is formed by $M \times 1$ blocks $\hv_{\ell,k}$, 
denoting the channel vectors between the $M$ antennas of RU $\ell$ and the single antenna of UE $k$.
For each $\ell \in \Lc, k \in \Kc$, the corresponding channel is a correlated complex circularly symmetric 
Gaussian vector with mean zero and 
covariance matrix 
\begin{align*}
\Sigmam_{\ell,k} = \EE[ \hv_{\ell,k}\hv_{\ell,k}^\herm],
\end{align*} where we use the well-known 
notation $\hv_{\ell,k} \sim \mathcal{CN} ({\bf 0},\Sigmam_{\ell,k})$.
The corresponding distance-dependent pathloss, also referred to as  large-scale fading coefficient (LSFC), is defined 
as 
\begin{align*}
\beta_{\ell,k} \defeq \frac{1}{M} \trace(\Sigmam_{\ell,k}),
\end{align*} and captures the average signal attenuation between 
RU $\ell$ and user $k$ due to distance and other macroscopic effects. Notice again that this model is completely standard
(e.g., see \cite{Noncooperative,Larsson-book,DemirFoundation,Subspace-Based} and references therein). 

\begin{figure}
	\centerline{
		\includegraphics[width=0.9\linewidth, trim={47 50 41 33}, clip]{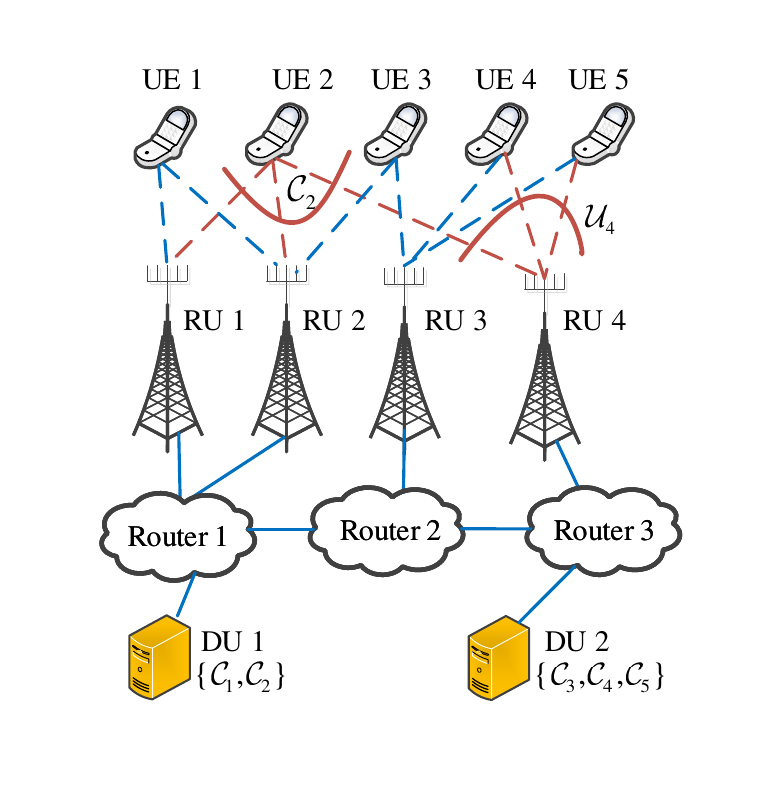} 
	}
	\caption{ An example network with $K = 5$ UEs, $L = 4$ RUs, $Q = 3$ routers, and $N = 2$ DUs.}
	\label{fig1}
\end{figure}

	{ 
	\begin{example} 
	In order to clarify the network topology and the localized processing in user-centric scalable CF-mMIMO networks, 
	consider the ``small network'' example given in Fig. \ref{fig1}, illustrating the data exchange required for cluster-level processing. Let us consider UE $2$ with user-centric cluster $\Cc_2$ 
	containing RUs $\{1, 2, 4\}$. In this example, the corresponding cluster processor is hosted by DU $1$. 
Therefore, in the UL, the RUs in $\Cc_2$ quantize their local observations relative to the UL signal sent by 
UE $2$ and send these quantized sample streams through the fronthaul to be conveyed to DU 1. 
In particular, the observations of RUs $\{1, 2\}$ are sent to router $1$ while 
RU $4$ forwards its observations to router $3$, which in turn forwards them to router $1$. 
All these observations are then sent directly from router $1$ to DU $1$ such that they can be jointly processed by the 
cluster processor for $\Cc_2$.  
Correspondingly, in the DL, the data for UE 2 shall be sent through the same routes in the reverse direction 
from DU $1$ to RUs $\{1, 2, 4\}$. \hfill $\lozenge$
	\end{example}
}

\subsection{Cluster Formation, Channel Estimation, and Local CSI}  \label{pilot-and-CSI}

In order to understand the decentralized scalable processing in both UL and DL, it is very important to consider how 
CSI is obtained at the RUs via pilot symbols transmitted by the users on the UL slot. 
By TDD reciprocity, the same CSI is also used for the DL multiuser MIMO precoding on the same resource block. 

We assume that UL pilot allocation and the formation of the user-centric clusters are performed jointly, and follow the subspace information aided overloaded pilot assignment  scheme in \cite{osawa2023overloaded}. 
All UEs transmit with the same average energy per symbol $P^{\rm ue}$ and we define the system parameter 
\begin{align*}
\SNR \defeq \frac{P^{\rm ue} }{N_0},
\end{align*} where $N_0$ denotes the complex baseband noise power spectral density.
When a user $k$ joins the system, it is associated to a cluster $\Cc_k$ of cardinality at most 
$\Cc_{\rm max}$ (a system parameter) of RUs with the largest LSFC, provided that 
\begin{gather}
	\beta_{\ell,k} \geq \frac{\eta}{M \SNR} , \label{eq:snr_threshold}
\end{gather}
where  $\eta > 0$ is a suitable threshold determining how much above the noise floor the useful signal in the presence of 
maximum possible beamforming gain (equal to $M$) should be. 
We assume that $\taudmrs$ out of $T$ signal dimensions per UL slot are dedicated to
pilots and use a codebook of $\taudmrs$ orthogonal pilot 
sequences $\{ \phiv_t : t \in [\taudmrs]\}$. 
Let $\Fm_{\ell,k}$ denote the matrix of the orthonormal eigenvectors of $\Sigmam_{\ell,k}$ spanning 
the {\em dominant channel subspace}, i.e., the subspace containing a sufficiently large fraction of the total channel energy 
$\trace(\Sigmam_{\ell,k}) = M \beta_{\ell,k}$ (e.g., see \cite{Subspace-Based,adhikary2013joint}). 
Then, we say that two users $k,k'$ are approximately mutually orthogonal in the spatial domain with respect to RU $\ell$ if 
\begin{align*}
\trace (\Fm_{\ell,k}^\herm \Fm_{\ell,k'} \Fm_{\ell,k'}^\herm \Fm_{\ell,k} ) \leq \epsilon,
\end{align*} for some fixed threshold $\epsilon$. 
In the association and pilot allocation scheme, a RU can assign a pilot to multiple UEs under the condition that
these users are approximately mutually orthogonal, according to the above definition.

For a given cluster association and pilot allocation, each RU $\ell$ obtains an estimate of the channel vectors 
$\{\hv_{\ell,k} : k \in \Uc_\ell\}$ using the subspace projection method of  \cite{Subspace-Based}. In particular, 
the pilot field received at RU $\ell$ is given by the $M \times \taudmrs$ matrix 
\begin{equation} 
	\Ym_{\ell}^{\rm pilot} = \sum_{i=1}^K \hv_{\ell,i} \phiv_{t_i}^\herm + \Zm_{\ell}^{\rm pilot}, \label{Y_pilot}
\end{equation}
where $\Zm_{\ell}^{\rm pilot}$ is additive white Gaussian noise (AWGN) 
with elements i.i.d. $\sim \Cc\Nc(0, 1)$, we use the pilot energy normalization
\begin{align*}
\| \phiv_{t} \|^2 = \taudmrs \SNR 
\end{align*} 
for all $t \in [\taudmrs]$, and $t_i$ denotes the index of the pilot sequence $\phiv_{t_i}$ allocated to user $i$. 
Then, RU $\ell$ projects the received pilot field onto a pilot sequence $t \in [\taudmrs]$, obtaining
\begin{eqnarray} 
	\yv^{(t)}_{\ell} & = & \frac{1}{\taudmrs \SNR} \Ym^{\rm pilot}_{\ell} \phiv_{t}  
	= \sum_{ i  : t_i = t} \hv_{\ell,i}  + \widetilde{\zv}^{(t)}_{\ell}.  \label{chest}
\end{eqnarray} 
Beyond the projected AWGN $\widetilde{\zv}^{(t)}_{\ell}$  with i.i.d. components  $\sim \Cc\Nc(0, \frac{1}{\taudmrs\SNR})$, 
$\yv^{(t)}_{\ell}$ contains the superposition of the channels for all users $i$ sharing the same pilot $t$. 
Let's focus on some specific user $k$ with $t_k = t$. 
With respect to such user,  the term 
$\sum_{ i \neq k : t_i = t} \hv_{\ell,i}$ is the so-called ``pilot contamination'' \cite{Noncooperative,Larsson-book,DemirFoundation,Subspace-Based}.
The CSI for user $k$ at RU $\ell$ is then achieved by the subspace projection
pilot decontamination scheme from \cite{Subspace-Based}, i.e., the orthogonal projection of 
$\yv^{(t)}_{\ell}$ onto the channel subspace of user $k$,  given by
\begin{align}
	\widehat{\hv}_{\ell,k} & = \Fm_{\ell,k}\Fm_{\ell,k}^\herm \yv^{(t)}_{\ell}  \notag
	 \\
	& = \hv_{\ell,k}  + \Fm_{\ell,k}\Fm_{\ell,k}^\herm \left ( \sum_{ i \neq k : t_i = t} \hv_{\ell,i} \right ) + \Fm_{\ell,k} \Fm_{\ell,k}^\herm \widetilde{\zv}^{(t)}_{\ell}.   \label{chest1}
\end{align}
Writing explicitly the pilot contamination term after the subspace projection, we obtain its covariance matrix as 
\begin{align*}
\Sigmam_{\ell,k}^{\rm co}  = \sum_{i \neq k : t_i = t} \Fm_{\ell,k} \Fm_{\ell,k}^\herm \Sigmam_{\ell,i} \Fm_{\ell,k} \Fm^\herm_{\ell,k} .
\end{align*}
When $\Fm_{\ell,k}$ and $\Fm_{\ell,i}$ are nearly mutually orthogonal, i.e. $\Fm_{\ell,k}^\herm \Fm_{l.i} \approx \zerov$,
the subspace projection is able to significantly reduce the pilot contamination effect. In most of the concurrent literature \cite{bjo2020,DemirFoundation,miretti2022}, the channel statistics (in particular, the covariance matrices $\Sigmam_{\ell,k}^{\rm co}$) are assumed to be known.
Schemes for channel subspace and covariance matrix estimation, respectively, in cell-free massive MIMO are presented in \cite{Subspace-Based, 9715152}. In particular, the scheme in 
\cite{Subspace-Based} is shown to achieve essentially the same performance of ideal channel subspace knowledge.
Hence, for simplicity, in this work we assume that the subspace information $\Fm_{\ell,k}$ for all $(\ell, k) \in \Ec_{\rm ran}$
is perfectly known, as justified by the results of \cite{DemirFoundation,Subspace-Based}.

Notice that each RU has only a partial (and noisy) CSI. The ``noisiness'' is given by the fact that the 
estimates in (\ref{chest1}) are affected by AWGN and residual pilot contamination. In addition, even in the absence of 
noise and pilot contamination, the CSI is partial in the sense that each RU $\ell$ can know only the channels
of users $k \in \Uc_\ell$. Because of the scalable decentralized user-centric cluster processing, even after gathering all the CSI collected at each RU $\ell \in \Cc_k$, the cluster processor of user $k$ (located in some DU) 
knows only the channels of RU-UE pairs $(\ell,k') \in \Ec_{\rm ran}$, where $\ell \in \Cc_k$ and $k' \in \Uc_{\Cc_k} \defeq \cup_{\ell \in \Cc_k} \Uc_\ell$. 
In general, this user subset is strictly smaller than the set of all users $\Kc$. 

\subsection{Uplink Signal Model with Fronthaul Quantization}   \label{uplink-and-quantization}

Let $\sss^{\rm ul} \in \CC^{K \times 1}$  denote the vector of modulation symbols transmitted collectively by the $K$ users over the UL at a given channel use of a generic resource block (since it is irrelevant, we omit also the channel use index).
The symbols $s_k^{\rm ul}$ ($k$-th components of $\sss^{\rm ul}$) are i.i.d. with mean zero and unit variance.  
The aggregate received signal of dimension $LM \times 1$ across all the RUs is given by 
\begin{equation}
	\yy^{\rm ul} = \sqrt{\SNR} \; \HH  \sss^{\rm ul} + \zz^{\rm ul}, \label{ULchannel-total}
\end{equation}
where $\zz$ is the AWGN with components $\sim \Cc\Nc(0,1)$ and 
$\yy = [(\yv_1^{\rm ul})^\transp, (\yv_2^{\rm ul})^\transp, \ldots, (\yv_L^{\rm ul})^\transp]^\transp$ 
is formed by $L$ blocks of dimension $M \times 1$. Each $\ell$-th block
\begin{equation} 
	\yv_\ell^{\rm ul} = \sqrt{\SNR}  \sum_{i\in \Kc} \hv_{\ell,i} s_i^{\rm ul} + \zv_\ell^{\rm ul} \label{ULchannel-RU}
\end{equation}
is the received signal at the $M$ antennas of the RUs $\ell$. 

Following \cite{DemirFoundation,Subspace-Based}, we consider 
``local detection with cluster-level combining''. In such scheme, for each UE $k \in \Kc$, each RU $\ell \in \Cc_k$
produces a local observation $r^{\rm ul}_{\ell,k}$ of symbol $s^{\rm ul}_k$ and sends it through the fronthaul 
network to the cluster processor of cluster $\Cc_k$, located in some DU. 
The local observation is obtained as a linear projection of $\yv_\ell^{\rm ul}$ onto a suitably defined receiver vector
$\vv_{\ell,k}$, such that 
\begin{equation}
	r^{\rm ul}_{\ell,k} = \vv_{\ell,k}^\herm \yv_\ell^{\rm ul} \; . \label{local-receiver}
\end{equation} 
The receiver vectors $\{\vv_{\ell,k} : k \in \Uc_\ell\}$ are computed at RU $\ell$ based on the local channel estimates
$\{\widehat{\hv}_{\ell,k} : k \in \Uc_\ell \}$. In this work, we consider the so-called ``plug-in'' linear MMSE receivers 
given the partial local CSI at RU $\ell$, while treating the unknown interference plus noise
term $\sqrt{\SNR} \sum_{i \notin \Uc_\ell} \hv_{\ell,i} s_i^{\rm ul} +  \zv_\ell^{\rm ul}$
as white Gaussian noise term \cite{goettsch2021impact} with variance per component given by 
\begin{equation} 
	\nu_{\ell} \eqdef  1 + \SNR \sum_{i \neq \Uc_{\ell}}  \beta_{\ell,i} \; . \label{sigmaell} 
\end{equation}
Under the assumption that the channels of all users $i \in \Uc_\ell$ are perfectly known and 
the noise plus out of cluster interference is a spatially white process with per-component variance $\nu_\ell$, 
the  linear MMSE receiving vector for user $k$ at RU $\ell \in \Cc_k$  is given by 
\begin{equation} 
	\vv_{\ell,k} = \left ( \nu_\ell \Id + \SNR \sum_{i \in \Uc_{\ell}} \widehat{\hv}_{\ell,i} \widehat{\hv}_{\ell,i}^\herm \right )^{-1} \widehat{\hv}_{\ell,k} \; .  \label{eq:lmmse}
\end{equation}
The cluster processor for user $k$ collects the local observations from all RUs $\ell \in \Cc_k$ and forms
a cluster-level combined observation which is then passed to  the channel decoder for user $k$ as the soft-output of a virtual single user additive noise channel.

Since the fronthaul network has finite capacity links, it is impossible to convey the samples continuous-valued 
$r^{\rm ul}_{\ell,k}$. Then, some form of quantization is needed at the RUs.  
In this work, we assume that each RU quantizes its local  observation $r^{\rm ul}_{\ell,k}$
with $B_{\ell,k}$ bits per sample, where $\{B_{\ell,k}\}$ are system parameters. 
For the sake of analytical tractability, instead of considering some specific quantization schemes, we make use of the 
following information theoretic results. 

\begin{definition} \label{defiQ} {\bf Lossy source coding with quadratic distortion.}
	Let $\{X_i\}$ denote a complex circularly symmetric stationary ergodic source with mean zero and variance $\sigma^2$.
	For any $n = 1, 2, \ldots, $ define the random vector 
	$X^n = (X_1, \ldots, X_n)$ extracted from the process $\{X_i\}$. A lossy source coding scheme of block length $n$ and rate $R$ bits/source symbol consists of an encoder $\phi_n : X^n \mapsto m$ that maps source blocks of length $n$ into quantization indices $m \in [1: 2^{nR}]$, and a decoder $\psi_n : m \rightarrow \widehat{X}^n$ 
	that maps quantization indices into the corresponding complex-valued representation vectors. 
	For given $(\phi_n, \psi_n)$, the achieved per-symbol mean-squared error (MSE)  distortion is defined as
	\[ D(\phi_n,\psi_n) = \frac{1}{n} \EE[ \| X^n - \psi_n(\phi_n(X^n))\|^2]. \]
	For each $D \geq 0$, the rate-distortion function $R_X(D)$ for the source $\{X_i\}$ under  MSE distortion is defined as \cite{Cover:2006}
	\begin{equation} 
		\begin{split}
			R_X(D) =  \inf \left \{ R : \exists \; \mbox{coding schemes $(\phi_n,\psi_n)$}  \right. \\ \left. \mbox{s.t. }
			\limsup_{n \rightarrow \infty} D(\phi_n,\psi_n) \leq D \right \}  \; .   \end{split}
	\end{equation} 
	$\phantom{x}$ \hfill $\lozenge$
\end{definition}

While the explicit expression of $R_X(D)$ of a general stationary ergodic source with finite variance 
depends on the source statistics and may be difficult to determine (in particular, in closed form), here we are interested in 
the following achievable upper bound, which is directly related to practical quantization. 

\begin{lemma}  \label{lemmaQ}
	Let $\{X_i\}$ denote a source as in Definition \ref{defiQ}. There exist lossy coding schemes for $\{X_i\}$ achieving MSE distortion $D$ at rate
	\begin{align*} 
	R = \max \{ \log_2 ( \frac{\sigma^2}{D} ), 0\}.
	\end{align*}
\end{lemma}

\begin{proof} 
	See Appendix \ref{AppendixQ}. 
\end{proof}

In order to obtain an analytically tractable expression of the quantized local observations, we use the following representation lemma to establish the relation between fronthaul rates and data rates in the wireless segment through the quantization at the RUs in the UL:

\begin{lemma} \label{lemmaR}
	Let $X_i$ and $\widehat{X}_i$ be the $i$-th source symbol and the corresponding $i$-th symbol of 
	the quantization codeword $\widehat{X}^n = \phi_n(X^n)$ for the schemes of Lemma \ref{lemmaQ}. 
	Then, we can write
	\begin{equation} 
		\widehat{X}_i = \alpha X_i + E_i,  \label{bussgang}
	\end{equation}
	where
	\begin{equation}
		\alpha = \frac{\sigma^2 - D}{\sigma^2}, \label{bussgangalpha}
	\end{equation}
	and where $E_i$ is a zero-mean random variable uncorrelated with $X_i$ with variance
	\begin{equation}
		\EE[|E_i|^2] = (1 - D/\sigma^2) D.
	\end{equation}
\end{lemma}

\begin{proof} 
	See Appendix \ref{AppendixQ}. 
\end{proof}

\begin{remark} {\bf On the practical achievability of the rate-distortion bound.}
	From \cite{ziv1985universal} we also know that the achievable quantization rate 
	$\max \left \{ \log_2 ( \frac{\sigma^2}{D}), 0 \right \}$ of Lemma \ref{lemmaQ} 
	can be achieved within an additive gap $\approx 1.5$ 
	bit/symbol by using entropy-coded dithered scalar quantization. 
	Furthermore, this gap reduces to $\approx 0.5$ bit/symbol in the high resolution regime, i.e., when $D \ll \sigma^2$. 
	Therefore, we shall treat the rate in Lemma \ref{lemmaQ} as ``practically achievable'' even without
	using high-dimensional vector quantization. 
	\hfill $\lozenge$
\end{remark}

For all $(\ell,k) \in \Ec_{\rm ran}$, define 
\begin{align*}
\sigma_{\ell,k}^2 \defeq \EE[ |r_{\ell,k}^{\rm ul}|^2]
\end{align*} to be the variance of 
the local observation (\ref{local-receiver}). For a given desired distortion level $D$, 
using Lemma \ref{lemmaQ}, for each $k \in \Kc$, each RU $\ell \in \Cc_k$ uses quantization rate
\begin{equation} 
	B_{\ell,k}  = \max \left \{ \log_2 \frac{\sigma_{\ell,k}^2}{D}, 0 \right \} \;  \label{rateQ}
\end{equation}
to send the local observation $r_{\ell,k}^{\rm ul}$ to the corresponding cluster processor.
If $\sigma^2_{\ell,k} < D$ for some $(\ell,k) \in \Ec_{\rm ran}$, 
then the local observation $r_{\ell,k}^{\rm ul}$ is simply not forwarded since in this case the quantization 
rate is $B_{\ell,k} = 0$. This is equivalent to removing RU $\ell$ from cluster $\Cc_k$. 
Hence, the user-centric cluster association does not depend only on the received signal strength or distance
	between users and RUs, but also on the target distortion level $D$ of the fronthaul quantization. 
	This effect will be further investigated in our numerical results of Section \ref{sec:simulation}.

From the proof of Lemma \ref{lemmaQ}, we know that there exist high-dimensional vector 
quantizers achieving MSE distortion $D$ with rate $B_{\ell,k}$, such that the joint marginal statistics 
between $r_{\ell,k}^{\rm ul}$ and its quantized 
representation $\widehat{r}_{\ell,k}^{\rm ul}$ is given by
\begin{align*} 
r_{\ell,k}^{\rm ul} = \widehat{r}_{\ell,k}^{\rm ul} + q_{\ell,k}
\end{align*} where $q_{\ell,k} \sim \Cc\Nc(0,D)$ and it is independent of 
$\widehat{r}_{\ell,k}^{\rm ul}$. Hence,  using Lemma \ref{lemmaR}, we can represent the quantized local observation as
\begin{equation} 
	\widehat{r}_{\ell,k}^{\rm ul} = \alpha_{\ell,k} r_{\ell,k}^{\rm ul}  + e_{\ell,k}, \label{bussgang_deco}
\end{equation}
where 
\begin{equation} 
	\alpha_{\ell,k} =  \frac{\sigma_{\ell,k}^2 - D}{\sigma_{\ell,k}^2}, \label{alpha_bussgang}
\end{equation}
and where $e_{\ell,k}$ is zero-mean variable uncorrelated with $r_{\ell,k}^{\rm ul}$ and with variance 
\begin{equation}
	\widehat{\sigma}^2_{\ell,k} = (1 - D/\sigma_{\ell,k}^2) D.  \label{error_bussgang}
\end{equation} 
In passing, we notice that  the representation (\ref{bussgang_deco}) for the quantized signal
can also be derived by the so-called {\em Bussgang decomposition} \cite{Demir_2021}, 
which has been used in several works to derive analytically tractable results for 
receivers with finite A/D conversion. 
	
	\begin{remark}  {\bf On the consistent definition of rates.}
		In this paper, we express both the PHY user rates and the fronthaul link load in 
		bit/channel use (equiv., bit/signal dimension, or bit/s/Hz).
		This is fully consistent since the PHY radio communication takes place on a channel bandwidth of $W$ Hz, 
		requiring a sampling rate of $W$ complex samples per second.  Hence, conveying the local observation 
		$\widehat{r}_{\ell,k}^{\rm ul}$ at $B_{\ell,k}$ bit per sample requires 
		a fronthaul load of $W B_{\ell,k}$ bit/second.  Also, the user PHY rate (in the UL or DL) of $R_k$ bit/s/Hz
		yields $W R_k$ bit/s. Hence, all rates are normalized by the channel bandwidth 
		and can be converted into ``practically meaningful''  bit/s simply by multiplying by $W$. \hfill $\lozenge$  \label{remark_rates}
	\end{remark}
	
	Letting $\widehat{\rv}_k^{\rm ul}$ the $|\Cc_k| \times 1$ vector of the quantized local observations
	$\{\widehat{r}^{\rm ul}_{\ell,k} : \ell \in \Cc_k\}$, the cluster processor produces the cluster-level combined observation 
	\begin{equation} 
		r_k^{\rm ul} = \wv_k^\herm \; \widehat{\rv}_k^{\rm ul}. \label{cluster-combining}
	\end{equation}
	The cluster-level combining coefficients $\wv_k$ are chosen to maximize the 
	``nominal'' Signal to Interference plus Noise Ratio (SINR) of the channel with input $s_k^{\rm ul}$ and output
	$r_k^{\rm ul}$, given the local knowledge of the cluster processor $\Cc_k$.~\footnote{Since each cluster processor has a partial channel knowledge, due to the local pilot-based channel state information as explained in Section \ref{pilot-and-CSI}, we distinguish between ``nominal'' and ``actual'' SINR. The optimization of the combining coefficients must consider the available channel state information, and therefore is based on the nominal SINR. In contrast, the actual achievable PHY user rates must be computed on the basis of the actual SINR, as it will be detailed later.}
	For this purpose, for all $\ell \in \Cc_k$,  let $\alpha_{\ell,k}$ be given from (\ref{alpha_bussgang})
	and define 
	\begin{align*}
	g_{\ell,k,i} \defeq \alpha_{\ell,k} \vv_{\ell,k}^\herm \widehat{\hv}_{\ell,i}
	\end{align*}
	for all $i \in \Uc_\ell$, the vector 
	\begin{align*}
	\av_k \eqdef \{ g_{\ell, k,k} : \ell \in \Cc_k\},
	\end{align*} and the matrix 
	$\Gm_k$ of dimension $|\Cc_k| \times ( | \Uc_{\Cc_k}| - 1)$ containing 
	in position corresponding to RU $\ell$ and UE $i \in \Uc_{\Cc_k} \setminus \{k\}$ (after a suitable index reordering)
	the coefficient $g_{\ell, k, i}$  if $(\ell,i) \in \Ec_{\rm ran}$, and zero elsewhere. Also, let 
	$\sv_k^{\rm ul}$ denote the $( | \Uc_{\Cc_k}| - 1) \times 1$ vector of the symbols of users $i \in 
	\Uc_{\Cc_k} \setminus \{k\}$. Then, from (\ref{local-receiver}) and (\ref{bussgang_deco}), the nominal model
	for the vector of quantized observations at the $k$-th cluster processor given the CSI known at the cluster processor for user $k$
	takes on the form
	\begin{equation}
		\widehat{\rv}^{\rm ul-nom}_k =  
		\sqrt{\SNR}   \left ( \av_k s^{\rm ul}_k  +  \Gm_k \sv_k^{\rm ul} \right ) + \zetav_k  + \ev_k,  \label{bussgang-vector-nominal}
	\end{equation} 
	where we define 
	\begin{align*}
	\ev_k \eqdef \{e_{\ell,k} : \ell \in \Cc_k\}
	\end{align*}   
	and the ($\alpha$-scaled) noise plus unknown interference vector $\zetav_k$ with elements 
	\[ \zeta_{\ell,k} = \alpha_{\ell,k} \vv_{\ell,k}^\herm \left ( \sqrt{\SNR} \sum_{i \notin \Uc_\ell} \hv_{\ell,i} s_i^{\rm ul} + \zv_\ell^{\rm ul} \right ), \]
	with variance 
	\begin{equation} 
		d_{\ell,k} \eqdef \EE[|\zeta_{k,\ell}|^2] = \alpha_{\ell,k}^2 \|\vv_{\ell,k}\|^2 \nu_\ell, \label{dellk}
	\end{equation}
	where $\nu_\ell$ is given in (\ref{sigmaell}). 
	
	The corresponding nominal SINR for UE $k$ with combining (\ref{cluster-combining}), based on 
	the nominal model (\ref{bussgang-vector-nominal}), is given by 
	\begin{equation} 
		\SINR^{\rm ul-nom}_k = \frac{\SNR \; \wv_k^\herm \av_k \av_k^\herm \wv_k}{\wv_k^\herm \Gammam_k  \wv_k},  \label{SINRnom}
	\end{equation}
	where we define 
	\begin{align*}
	\Gammam_k \defeq \SNR \; \Gm_k \Gm^\herm_k + \diag(\{d_{\ell,k} + \widehat{\sigma}_{\ell,k}^2: \ell \in \Cc_k\}).
	\end{align*}
	 The maximization of (\ref{SINRnom})
	with respect to the vector of combining coefficients $\wv_k$ is a classical generalized Rayleigh quotient maximization 
	problem, solved by finding the generalized eigenvector of the maximum generalized eigenvalue of the matrix pencil 
	$(\av_k \av_k^\herm, \Gammam_k )$.  Since the matrix $\av_k \av_k^\herm$ has rank 1, the solution is immediately given by  
	\begin{align}
		\wv_k = \Gammam_k^{-1} \av_k. \label{eq:optimal_wk}
	\end{align}
	Notice that \eqref{eq:optimal_wk} reduces to that in \cite{Subspace-Based} in the high quantization resolution limit 
	(i.e.,  $D \rightarrow 0$). 
	
	\begin{remark} {\bf Computability of the quantization rates.}
		The quantization rates $B_{\ell,k}$, as well the coefficients $\alpha_{\ell,k}$ and the 
		quantization error variances $\widehat{\sigma}^2_{\ell,k}$ are determined by the 
		variance of the local observations $\sigma^2_{\ell,k}$ at the RUs. 
		This parameter is an ensemble average, or equivalently, an ergodic time-average over a long sequence of 
		resource blocks, since the fronthaul quantization rate allocation should depend on long-term average quantities, and not on 
		the instantaneous realizations that change on a block by block basis. 
		Nevertheless, because of the intricate dependency between the local receiver vector $\vv_{\ell,k}$, 
		the channel estimates $\{\widehat{\hv}_{\ell,i} : i \in \Uc_\ell\}$ and the true channel vectors, it is not generally possible to give 
		a simple closed form. In our numerical results, for a given network configuration, we have computed 
		$\sigma^2_{\ell,k} = \EE[|r^{\rm ul}_{\ell,k}|^2]$ by Monte Carlo expectation. In practice, measuring the average signal power at the output of the local estimator as a time-average can be easily done. As a matter of fact, this type of parameters are routinely computed 
		for the adaptive dynamic range of the A/D converter front-end. \hfill $\lozenge$
	\end{remark}
	
\begin{figure}
		\centerline{
			\includegraphics[width=0.9\linewidth, trim={5 5 5 5}, clip]{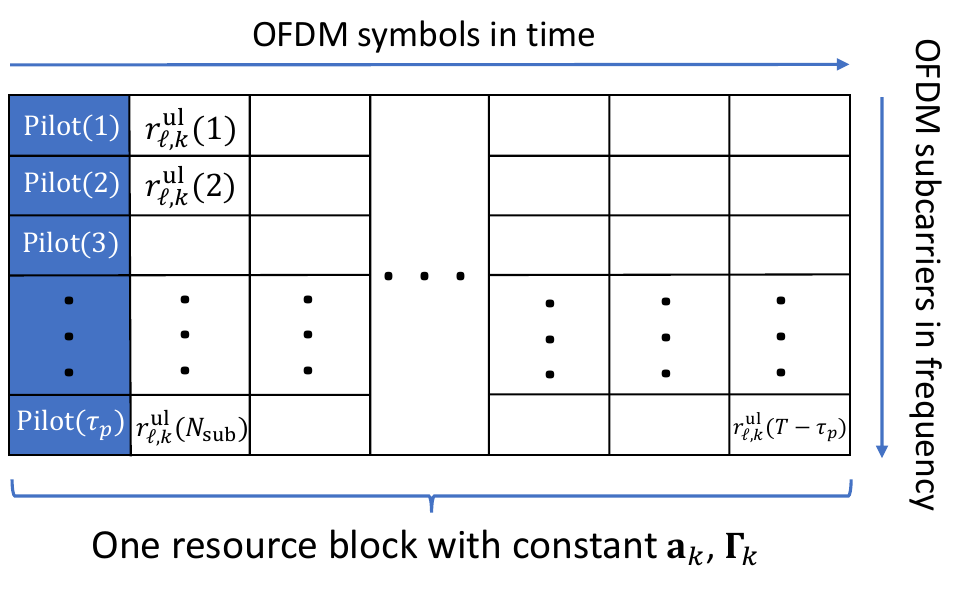} 
		}
		\caption{ Example of the data in a resource block with $T$ signal dimensions, where blue and white boxes represent UL pilot and data signals, respectively.}
		\label{coherence_block}
	\end{figure}

	\begin{remark} {\bf Quantized fronthaul vs. unquantized RU processing.}
		At this point, a reader may wonder why we have treated the processing in the RUs (i.e., UL channel estimation 
		and computation of the local receiver vectors $\vv_{\ell,k}$) as unquantized, while we have imposed quantization of
		the local measurements $r^{\rm ul}_{\ell,k}$. The reason is that the actual baseband front-end of the RU can be implemented
		with a finite arithmetic of sufficient resolution such that the local operations can be considered unquantized. 
		In contrast, the samples $r^{\rm ul}_{\ell,k}$ must be sent to the cluster processor $\Cc_k$ hosted 
		in some DU through the fronthaul. This represents the bulk of the UL data transmission in the fronthaul, since it costs $B_{\ell,k}$ bits per
		complex sample, at rate of $W$ complex samples per second ($W$ indicating the channel bandwidth). { The vector $\av_k$ and the matrix $\Gammam_k$ can be formed directly from the quantities received by the DU hosting cluster $\Cc_k$ because they depend only on local CSI, available at the RU. Please refer to the entries of $\Gm_k$ and $\av_k$ given above \eqref{bussgang-vector-nominal}, and note that all required quantities are available at the RUs and can be forwarded to the DU. The same holds for $\{d_{\ell,k}, \alpha_{\ell,k}, \widehat{\sigma}^2_{\ell,k}, \nu_{\ell} : \ell \in \Cc_k\}$.}
		Also, for simplicity we assumed that the coefficients in $\av_k$ and $\Gammam_k$ necessary to compute the cluster-level combining coefficients in (\ref{eq:optimal_wk}) are forwarded with sufficiently high resolution.
		
		{ We plot Fig. \ref{coherence_block} to illustrate the structure of a resource block with $T$ signal dimensions (channel uses), among which $\tau_p$ are used by the UL pilot signal for channel estimation. The remaining $T - \tau_p$ channel uses are used for data signals, where $r_{\ell,k}$ varies for each channel use, while $\av_k$ and $\Gammam_k$ remain \textit{constant} for all $T$ channel uses. 
		Assuming $\tau_p \approx 20$ and $T$ typically between $200$ and $2000$, 
		the number of observations per resource block, i.e., $ \{ r_{\ell,k}^{\rm ul}(t) : t \in [T-\tau_p] \} $, is relatively large compared to the dimensions of $\av_k \in \CC^{ | \Cc_k | \times 1 }$ and the (sparse) matrix 
		$\Gm_k \in \CC^{ |\Cc_k| \times ( | \Uc_{\Cc_k}| - 1) }$ used to compute $\Gammam_k$.
		
		This is why we can assume that $\av_k$ and $\Gm_k$ are forwarded with sufficiently high resolution to be considered unquantized, and why we need to quantize the $T - \tau_p$  samples of $r_{\ell,k}^{\rm ul}$. Notice that $\av_k$ and $\Gm_k$  depend only on local CSI. In the 5G standard, the coefficients required to compute $\av_k$ and $\Gm_k$ are collected by letting the users periodically send sounding reference signals (SRS) in the UL \cite{3gpp38211}. This means that a set of coefficients must be updated at each SRS period, which may very well span hundreds or thousands of time-frequency symbols (approximately of the same order of magnitude as $T$).}
		
		\hfill $\lozenge$
	\end{remark}
	
	We conclude this section by detailing the actual SINR and the resulting PHY per-user UL rate. 
	Writing explicitly the expression of $r^{\rm ul}_k$ from 
	(\ref{ULchannel-total}), (\ref{local-receiver}), (\ref{bussgang_deco}), and (\ref{cluster-combining}),
	we obtain~\footnote{Note that this expression is different from what obtained by the nominal model 
		(\ref{bussgang-vector-nominal}). However, while (\ref{bussgang-vector-nominal}) can be computed by the 
		$k$-th cluster processor,  (\ref{suca}) cannot, since it contains some unknown variables.} 
	\begin{eqnarray}
		r^{\rm ul}_k  & = & \sum_{\ell \in \Cc_k} w_{\ell,k}^* \widehat{r}^{\rm ul}_{\ell,k} \label{suca} \\
		& = & \sqrt{\SNR} \left ( \sum_{\ell \in \Cc_k} w_{\ell,k}^* \alpha_{\ell,k} \vv_{\ell,k}^\herm \hv_{\ell,k} \right ) s_k^{\rm ul} \label{useful-term} \\
		& & + \sqrt{\SNR} \sum_{i \neq k} \left ( \sum_{\ell \in \Cc_k} w^*_{\ell,k} \alpha_{\ell,k} \vv_{\ell,k}^\herm \hv_{\ell,i} \right ) s^{\rm ul}_i \label{multiuser-interference} \\
		& & +  \sum_{\ell \in \Cc_k} w^*_{\ell,k} \left ( \alpha_{\ell,k} \vv_{\ell,k}^\herm \zv_\ell^{\rm ul} + e_{\ell,k} \right ), 
		\label{channel-and-quantization-noise}
	\end{eqnarray}
	where (\ref{useful-term}) is the useful signal term, (\ref{multiuser-interference}) is the multiuser interference term, 
	and (\ref{channel-and-quantization-noise}) is the channel and quantization noise term.
	The resulting {\em actual} SINR conditioned on the realization of all the channel vectors, is given by 
	 \begin{equation} 
		\SINR^{\rm ul}_k  = \frac{  \SNR \left |  \sum_{\ell \in \Cc_k} \widetilde{g}_{\ell, k, k} \right |^2 }
		{\sum_{\ell \in \Cc_k}  \widetilde{d}_{\ell,k}
			+ \SNR \sum_{i \neq k} \left |  \sum_{\ell \in \Cc_k} \widetilde{g}_{\ell, k, i}  \right |^2 } ,  \nonumber
	\end{equation}
	where 
	\begin{align*}
	\widetilde{g}_{\ell, k, i} \eqdef w^*_{\ell,k} \alpha_{\ell,k} \vv_{\ell,k}^\herm \hv_{\ell,i}
	\end{align*} 
	\begin{align*}
	\text{and \ } \widetilde{d}_{\ell,k} \eqdef |w_{\ell,k}|^2 \left ( \alpha_{\ell,k}^2 \|\vv_{\ell,k}\|^2 + \widehat{\sigma}^2_{\ell,k} \right ).
	\end{align*}
	As a performance measure of the UL physical layer,  we consider the so-called {\em Optimistic Ergodic Rate} (OER) 
	\cite{8304782} given by 
	\begin{eqnarray}
		R^{\rm ul}_k = \EE [ \log (1 + \SINR^{\rm ul}_k) ], \label{ergodic_rate_ul}
	\end{eqnarray}
	where the expectation is with respect to the small-scale fading, for given 
	values of the LSFCs that depend on the placement of UEs and RUs on the plane, distance-dependent pathloss function, 
	and cluster formation. An extensive motivation for using OER as performance metric is provided in \cite{Subspace-Based} 
	and will not be repeated here for the sake of space limitation. 
	
	\subsection{Downlink Transmission} \label{DL-transmission-and-precoding}
	
	We denote by $\xx$ the precoded signal vector of dimension $LM \times 1$ transmitted simultaneously 
	by all RUs in one downlink channel use. Using linear precoding, this is given by 
	$\xx = \UU \sss^{\rm dl}$, where 
	$\sss^{\rm dl} \in \CC^{K \times 1}$ is the vector of (coded) information symbols for the $K$ users 
	(independent with mean zero and variance $q_k \geq 0$), and 
	$\UU \in \CC^{LM \times K}$ is the overall DL precoding matrix with unit-norm columns.
	In order to preserve the per-cluster processing condition, this is a block matrix formed by $M \times 1$ blocks $\uv_{\ell,k}$ such that 
	$\uv_{\ell,k} = \zerov$ if $(\ell, k) \notin \Ec_{\rm ran}$. 
	Notice that $\xx$ can be computed by local computations at the RUs as follows. For each $\ell \in \Lc$, RU $\ell$ computes
	\begin{equation} 
		\xv_\ell = \sum_{k \in \Uc_\ell} \uv_{\ell,k} s_k^{\rm dl} \; . \label{local-tx-vector}
	\end{equation}
	Then, $\xx = [\xv^\transp_1, \xv^\transp_2, \ldots, \xv^\transp_L]^\transp$. In order to compute
	(\ref{local-tx-vector}), RU $\ell$ needs the information symbols $s_k^{\rm dl}$ for all users $k \in \Uc_\ell$, and the corresponding
	local precoding vectors $\uv_{\ell,k}$. Again, we notice that the precoding vectors are constant for the whole DL slot of $T$ channel uses, since they depend on the CSI.  Therefore, the fronthaul load incurred by serving user $k$ in the DL consists of 
	conveying the information bits to produce the information symbols $s^{\rm dl}_k$ to each RU $\ell \in \Cc_k$.  For the DL, the most efficient form of ``fronthaul compression'' is provided by simply sending 
the information bits from the DUs to the RUs, and letting the RUs encode, modulate, and precode, the signals to be sent to the UEs via the DL. This is most efficient because the information bits are incompressible by definition. Hence, no more parsimonious representation of the DL signals can be obtained. 
It should be noticed that the local combining and quantization of the per-UE UL data streams
is fully compatible with the 7.2 ``split'' in the specification of disaggregated architecture in 3GPP recommendations. 
In contrast, sending the information bits to the RUs and letting the RUs perform encoding, modulation, and precoding, 
is (approximately) compatible with the 7.3 ``split'' of 3GPP (e.g., see \cite{sup6620205g}).
	
	For the downlink precoding vectors, we use the approximate UL-DL reciprocity of \cite{Subspace-Based} and let
	$\uv_{\ell,k} \propto w^0_{\ell,k} \vv_{\ell,k}$ where $\vv_{\ell,k}$ are the local linear MMSE combiners for the UL defined in (\ref{eq:lmmse}) and $w^0_{\ell,k}$ are the cluster-level combining coefficients defined in (\ref{eq:optimal_wk}) for the case of 
	zero quantization distortion, i.e., replacing $\alpha_{\ell,k} = 1$ and $\widehat{\sigma}^2_{\ell,k} = 0$ in the expressions for
	$\av_k$ and $\Gammam_k$ in (\ref{eq:optimal_wk}). Then, the actual local precoding vectors $\uu_k$ are obtained by 
	stacking the blocks $w^0_{\ell,k} \vv_{\ell,k}$ for all $\ell \in \Cc_k$ and all-zero blocks for $\ell \neq \Cc_k$, and normalizing the 
	resulting vector of dimension $LM \times 1$ to have unit norm. Notice that in this way, apart the common normalization factor, 
	$\uv_{\ell,k}$ can be calculated from the CSI present at RU $\ell$, with sufficiently high resolution finite arithmetic. 
	
	Letting $\hh_k$ denote the $k$-th column of the overall channel matrix $\HH$, 
	the received signal sample at the (single-antenna) port of UE $k$ corresponding to one DL channel use is given by 
	\begin{eqnarray} 
		y_k^{\rm dl} & = & \hh_k^\herm \xx  + z_k^{\rm dl} \nonumber \\
		& = & \hh_k^\herm \uu_k s^{\rm dl}_k   + \sum_{j \neq k} \hh_k^\herm \uu_j s^{\rm dl}_j  + z^{\rm dl}_k , 
		\label{DLchannel}
	\end{eqnarray}
	where, without loss of generality, we scale the received signal such that the 
	noise is $z_k^{\rm dl} \sim \Cc\Nc(0, \SNR^{-1})$. 
	The corresponding DL SINR is given by 
	\begin{eqnarray}
		\SINR^{\rm dl}_k & = & \frac{|\hh_k^\herm \uu_k|^2 q_k}{\SNR^{-1} + \sum_{j\neq k}   |\hh_k^\herm \uu_j|^2 q_j }. \label{DL-SINR} 
	\end{eqnarray}
	Since $\UU$ has unit-norm columns, we obtain  
	\begin{align*}
	\trace \left ( \EE [ \xx \xx^\herm ] \right )  =  \sum_{k=1}^K q_k.
	\end{align*}
	Imposing that the total transmission powers in the UL and DL are both equal to $K P^{\rm ue}$, with the above normalizations, we obtain the condition $\sum_{k \in \Kc} q_k = K$ for the power coefficients of individual user DL data streams.
	For simplicity, in this paper we choose uniform power allocation to all data streams, i.e., $q_k = 1$ for all $k \in \Kc$ (see also \cite{Subspace-Based,buzzi2017downlink}). 
	The corresponding OER for the DL is given by 
	\begin{eqnarray}
		R^{\rm dl}_k = \EE [ \log (1 + \SINR^{\rm dl}_k) ]. \label{ergodic_rate_dl}
	\end{eqnarray}
	
	\begin{remark} {\bf Multiple-multicast nature of the DL fronthaul traffic.}
		As said before, the number of information bits per channel use necessary to encode the DL signal 
		for user $k$ at each RU $\ell \in \Cc_k$ is equal to $R^{\rm dl}_k$. Notice that the DL fronthaul traffic is of multiple multicast type, since the same information stream of user $k$, originated at the cluster processor of user $k$ hosted in some DU, 
		must be conveyed to all RUs $\ell \in \Cc_k$, for all $k \in \Kc$.  \hfill $\lozenge$ \label{remark_dl_fronthaul}
	\end{remark}

	\section{Joint Fronthaul Load Balancing and Computation Resource Allocation}

\label{sec:UL-Distributed}

In this section, we describe the fronthaul load balancing and cluster placement problem. 
We aim to minimize the maximum load over all links in the fronthaul network, described by the graph $\Gc_{\rm front}(\Lc, \Qc, \Nc, \Ec _{\rm front})$.
We first formulate the problem constraints for the UL and the DL separately. 
	Then, we will consider two alternative approaches:  1) 
	joint optimization with full-duplex fronthaul; 2) joint optimization with half-duplex fronthaul.
The amount of UL fronthaul load transmitted from RU $\ell$ to router $q$, from router $q$ to router $q'$, and from 
router $q$ to DU $n$ regarding user $k$ is denoted 
as $x_{k}^{\text{ru}}(\ell, q)$, $x_{k}^{\text{fh}}(q,q')$, and $x_{k}^{\text{du}}(q,n)$, respectively. For the DL, we use a similar notation, where $y_{k}^{\text{ru}}(q, \ell)$, $y_{k}^{\text{fh}}(q,q')$ and $y_{k}^{\text{du}}(n, q)$ is the amount of DL fronthaul data with respect to user $k$ sent from router $q$ to RU $\ell$, from router $q$ to router $q'$, and from DU $n$ to router $q$, respectively. Note that the UE-RU associations are not part of the optimization problem. They are defined according to the cluster formation process described in Section \ref{pilot-and-CSI}.

\subsection{UL fronthaul load flow}

Recall that RU $\ell \in \Cc_k$ generates a quantization rate 
$\fhratekUL$ bit/symbol (see  \eqref{rateQ}) that must be conveyed via the fronthaul to DU $n$ hosting~\footnote{We say that DU $n$ ``hosts''  cluster $\Cc_k$ to indicate that it hosts the processor of 
	the user-centric cluster $\Cc_k$.} 
$\Cc_k$. 
In short, RUs represent multiple data nodes (each RU $\ell$ corresponds to $|\Uc_\ell|$ sources of rate $\fhratekUL$ for all $k \in \Uc_\ell$) and the DU are multiple sink nodes, where each DU $n$ acts as sink for all flows relative to the 
users $k$ whose cluster processor is hosted by DU $n$. 
Routers are intermediate relay nodes (not sources nor sinks). Since the UL fronthaul contains only unicast flows, 
routers must obey to the law of conservation of flow. In particular, for each user $k$, every router $q$ must satisfy~\footnote{It is understood that all 
	sums extend over existing links in  $\Ec _{\rm front}$, although this is not explicitly indicated.}
\begin{align} \label{eq}
	\sum_{\ell}x_{k}^{\text{ru}}(\ell, q)+\sum_{q''}x_{k}^{\text{fh}}(q'',q)   =   \sum_{n} x_{k}^{\text{du}}(q,n)	+\sum_{q'}x_{k}^{\text{fh}}(q,q') .
\end{align}
The load of any fronthaul link $(a,b) \in \Ec_{\rm front}$ is given by $\sum_{k} x_{k}(a,b)$, where 
$x_{k}(a,b)$ is one of the load variables previously introduced, depending of the nature of the link. 
The source constraint relative to RU $\ell$ and user $k$ is expressed by 
\begin{align}
	\sum_{q} x_{k}^{\text{ru}}(\ell, q) \geq a_{\ell, k}\fhratekUL , \; \forall k,\ell , \label{eq_ul_k_to_q_sum}
\\
\text{and }	 x_{k}^{\text{ru}}(\ell, q) \leq a_{\ell, k} \fhratekUL , \; \forall k,\ell,q,   \label{eq_ul_k_to_q}
\end{align}
where the UE-RU association binary variable $a_{\ell, k}$ is defined as
\begin{align}
	a_{\ell, k} = \begin{cases}
		1,& \text{if } (\ell, k) \in \Ec_{\rm ran}, \\
		0 ,& \text{if } (\ell, k) \notin \Ec_{\rm ran} . 
	\end{cases}   \nonumber
\end{align}
Now, let $b_{k,n} \in \{0,1\}$ denote the cluster processor placement variable, defined by 
\begin{align}
	b_{k, n} = \begin{cases}
		1 ,& \text{if  $\Cc_k$ is hosted by DU $n$},  \\
		0 ,&  \text{otherwise}.  
	\end{cases}   \nonumber
\end{align}
Since each $\Cc_k$ must be hosted by exactly one DU, we have the constraint
\begin{align}
	\sum_{n=1}^{N} b_{k,n}  =  1 , \; \forall k.  \label{sumb}
\end{align}
{ We impose the computation capacity constraint at DU $n$ given by
\begin{gather}
	\sum_{k = 1}^K b_{k,n} \leq Z_n,  \; \forall n ,
\end{gather}
where $Z_n$ is a computation limit for the number of cluster processors that can be hosted by any DU $n$. Notice that in order to have a feasible problem it must be $\sum_n Z_n  \geq K$.} Given the cluster processor placement, the sink constraint is obtained by noticing that the received flow relative to 
user $k$ to  DU $n$ hosting $\Cc_k$ must be not smaller than the sum of source rates 
$\fhratekUL$ over all RUs $\ell \in \Cc_k$, i.e., 
\begin{align}\label{d3}
	\sum_{q} x_{k}^{\text{du}}(q,n)\geq b_{k,n} \sum_{\ell \in \Cc_k} \fhratekUL , \; \forall k, n .
\end{align}
Although not strictly necessary for the minimization of the max link load, we also introduce the 
individual load variables upper bounds
\begin{align}\label{d4}
	x_{k}^{\text{du}}(q,n)\leq b_{k,n} \sum_{\ell \in \Cc_k}  \fhratekUL , \; \forall k, q, n .
\end{align}
In particular, this means that if $b_{k,n} = 0$ (i.e., DU $n$ does not host $\Cc_k$) the rate
relative to user $k$ from any connected router $q$ to $n$ will be zero. 

\subsection{DL fronthaul load flow}

As already anticipated in Section \ref{DL-transmission-and-precoding}, the DL is quite different from the UL regarding the fronthaul flow. The DL fronthaul traffic is {\em multiple multicast}. In fact, the data sent by DU $n$ 
regarding user $k$ is the same for all RUs $\ell \in \Cc_k$.
As explained in Remark \ref{remark_dl_fronthaul}, the number of information bits per channel use necessary 
to encode the DL signal for user $k$ at each RU $\ell \in \Cc_k$ is equal to the DL PHY rate $\fhratekDL$, and data can only be sent between nodes that are connected (i.e., with the edge topology defined 
by $\Ec_{\rm front}$). 

Because the DL fronthaul traffic is of multicast type, the flow conservation at the routers no longer applies (e.g., 
intermediate nodes may duplicate some input to multiple output links). 
Instead, considering the data of user $k$, the output data of a router $q$ to any RU $\ell$ or router $q''$ must be less than or equal to the sum input data from DUs and other routers, i.e.,
\begin{align}\label{dl1}
	\sum_{n} y_{k}^{\text{du}}(n,q)+\sum_{q'}y_{k}^{\text{fh}}(q',q)\geq y_{k}^{\text{ru}}(q,\ell), \; \forall k, q, \ell, 
\end{align}
and 
\begin{align}\label{dl2}
	\sum_{n} y_{k}^{\text{du}}(n,q)+\sum_{q'}y_{k}^{\text{fh}}(q',q)\geq y_{k}^{\text{fh}}(q,q''), \; \forall k, q, q'' .
\end{align}
In the DL, the DUs are (multiple) source nodes, and the RUs are (multiple) sink nodes. 
The DU hosting $\Cc_k$ must transmit at least $\fhratekDL$ bits/symbol to the connected routers, i.e., 
\begin{align}\label{dl3}
	\sum_{q} y^{\text{du}}_{k}(n,q) \geq b_{k,n} \fhratekDL, \; \forall k,n.
\end{align}
On each individual link to a router $q$, the DU needs to transmit at most $\fhratekDL$ bits and only the DU hosting cluster $\Cc_k$  transmits fronthaul data for user $k$. These two constraints are summarized as 
\begin{align}\label{dl4}
	y^{\text{du}}_{k}(n,q) \leq b_{k,n} \fhratekDL, \; \forall k,n,q.
\end{align}
The constraint that guarantees that each RU $\ell \in \Cc_k$ receives $\fhratekDL$ fronthaul bits for user $k$ is formulated as
\begin{align}\label{dl5}
	\sum_{q} y_{k}^{\text{ru}}(q, \ell) \geq a_{k,\ell}\fhratekDL, \; \forall k,\ell.
\end{align}

	\subsection{Full- and Half-Duplex Fronthaul Optimization Problems}   \label{jointopt}
	
	We formulate the joint optimization problem for an arbitrary DL/UL TDD resource allocation fraction 
	${ {\gamma_{\rm DL}}} \in (0,1)$. 
	We start by considering the case of full-duplex fronthaul. Notice that although in this case
	the constraints on the UL and DL traffic are separate, the optimization of UL and DL 
	must be performed jointly since the two problems are tied together by the common 
	cluster processor allocation variables $\Bc = \{ b_{k,n}\}$. 
	
	Let $\Cc = \{C_L^{\rm ul}, C_Q^{\rm ul}, C_D^{\rm ul},  C_L^{\rm dl}, C_Q^{\rm dl}, C_D^{\rm dl}\}$ the
	maximum link loads for UL and DL for  RU-router, router-router, and router-DU links (with corresponding weights $\eta_{L/Q/D}^{\rm ul/dl}$).
	Let $\Xc$ and $\Yc$ denote the ensemble of all UL load and DL load variables, respectively. 
	Then, the proposed joint optimization problem for the full-duplex fronthaul case is given by
	\begin{subequations}\label{opt_problem_fd}
		\begin{eqnarray}
			&\min\limits_{\Bc, \Cc, \Xc, \Yc} & \max \Big \{ (1 - { {\gamma_{\rm DL}}}) ( \eta_L^{\rm ul} C_L^{\rm ul}+ \eta_Q^{\rm ul} C_Q^{\rm ul}+ \eta_D^{\rm ul} C_D^{\rm ul}), \nonumber \\
			& & \; \;\;\;\;\;\;\;\;\;\; { {\gamma_{\rm DL}}} ( \eta_L^{\rm dl} C_L^{\rm dl}+ \eta_Q^{\rm dl} C_Q^{\rm dl}+ \eta_D^{\rm dl} C_D^{\rm dl}) \Big \} \\
			&\text{s. t.}
			&\sum_{k} x^{\text{ru}}_{k}(\ell, q) \leq C_L^{\rm ul}, \; \forall \ell, q, \label{FD1} \\
			&&\sum_{k}x_{k}^{\text{fh}}(q,q') \leq C_Q^{\rm ul}, \; \forall q, q',\\
			&&\sum_{k} x_{k}^{\text{du}}(q,n) \leq C_D^{\rm ul}, \; \forall q, n,\\
			&&\sum_{k}y_{k}^{\text{ru}}(q,\ell)\leq C_L^{\rm dl} , \; \forall q,\ell, \\
			&&\sum_{k}y_{k}^{\text{fh}}(q,q')\leq C_Q^{\rm dl} , \; \forall q, q',\\
			&&\sum_{k}y_{k}^{\text{du}}(n,q)\leq C_D^{\rm dl} , \; \forall n, q, \label{FD2}\\
			&&  \eqref{eq}-\eqref{dl5},
		\end{eqnarray}
	\end{subequations}
	where \eqref{FD1}-\eqref{FD2} ensure that the data flow transmitted within the UL/DL network does not exceed the respective maximum link capacity. 
	Next, we consider the case of half-duplex fronthaul.  
	In this case,  the flow constraints must incorporate the fact that 
	the data rate generated by the RUs in the UL is weighted by a factor $(1 - { {\gamma_{\rm DL}}})$ and the data rate
	generated by the DUs in the DL is weighted by a factor ${ {\gamma_{\rm DL}}}$. 
	In particular, the UL source constraints \eqref{eq_ul_k_to_q_sum}, \eqref{eq_ul_k_to_q} and the sink constraints
	\eqref{d3}, \eqref{d4} are modified by replacing  $B_{\ell,k}$ with 
	\begin{align*}
	\widetilde{B}_{\ell,k} = (1 - { {\gamma_{\rm DL}}}) B_{\ell,k}.
	\end{align*} Similar, for the DL, the source constraints \eqref{dl3} and the sink constraints \eqref{dl4}, \eqref{dl5}
	are modified by replacing  $R^{\rm dl}_{k}$ with 
	\begin{align*}
	\widetilde{R}^{\rm dl}_{k} = { {\gamma_{\rm DL}}} R^{\rm dl}_{k}.
	\end{align*}  
	Notice also that the flow conservation equation \eqref{eq} and the multicast routing inequalities \eqref{dl1}, \eqref{dl2}
	remain unchanged. For brevity, we shall refer to the ensemble of the above constraint equations 
	(modified and unchanged) for the half-duplex case as  ``HD-case constraints''. 
	
	Consider weights $\eta_L$, $\eta_Q$, $\eta_D$ and $\Cc^{\rm hd} = \left\{ C_L, C_Q, C_D \right\}$. Then, 
	the joint optimization problem for half-duplex fronthaul is given by 
	\begin{subequations}\label{opt_problem_hd}
		\begin{eqnarray}
			&\min\limits_{\Bc, \Cc^{\rm hd}, \Xc, \Yc} & \eta_L C_L + \eta_Q C_Q + \eta_D C_D \hspace{3.5cm} \\
			&\text{s. t.}
			& \sum_{k}  \left( x^{\text{ru}}_{k}(\ell, q) + y_{k}^{\text{ru}}(q,\ell)  \right)		\leq C_L, \; \forall \ell, q, \label{HD1}\\
			&&\sum_{k}  \left( x_{k}^{\text{fh}}(q,q') + y_{k}^{\text{fh}}(q,q')  \right)			\leq C_Q, \; \forall q, q',  \\
			&&\sum_{k}  \left( x_{k}^{\text{du}}(q,n) + y_{k}^{\text{du}}(n,q)  \right)		\leq C_D, \; \forall q, n, \label{HD2}\\	
			&&  \mbox{HD-case constraints,}
		\end{eqnarray}
	\end{subequations}
	where constraints \eqref{HD1}-\eqref{HD2} ensure that the sum data are less than the link capacity.
	Problems \eqref{opt_problem_fd} and \eqref{opt_problem_hd} are MILPs and can be solved by 
	existing standard and highly efficient optimization tools \cite{elie2019}. { To implement \eqref{opt_problem_fd} and \eqref{opt_problem_hd} in our simulations, we utilize the MATLAB function \texttt{intlinprog}. In practical systems, the joint load balancing and cluster processor placement problems (i.e., \eqref{opt_problem_fd} or \eqref{opt_problem_hd}) are typically handled by a network orchestrator with knowledge of the network statistics and topology/computation capacity constraints. This is usually implemented in software, operating on some central controller (e.g., in a cloud server \cite{de2019network}).}

\section{Numerical Results} \label{sec:simulation}

We consider a square with  area of $A = 200 \times 200\;\; {\rm m}^2$ with a torus topology to avoid boundary effects. 
The LSFCs (including distance-dependent pathloss, blocking effects, and shadowing) are given by the 3GPP urban microcell street canyon pathloss model from \cite[Table 7.4.1-1]{3gpp38901}, which differentiates between UEs in line-of-sight (LOS) and non-LOS (NLOS). The probability of LOS is distance-dependent and given in \cite[Table 7.4.2-1]{3gpp38901}. 
A log-normal Gaussian random variable with different parameters for LOS and NLOS is added to the deterministic pathloss model to account for shadow fading.

For the spatial correlation between the channel antenna coefficients, we consider a simple directional channel model defined as follows. 
The angular support 
\begin{align*}
\Theta_{\ell,k} = [\theta_{\ell,k} - \Delta/2, \theta_{\ell,k} + \Delta/2]
\end{align*}
 is centered at angle $\theta_{\ell,k}$ of the LOS
between RU $\ell$ and UE $k$ (with respect to the RU boresight direction), with angular spread $\Delta$. 
Additionally, let
\begin{align*} 
\hv_{\ell,k} = \sqrt{\frac{\beta_{\ell,k} M}{|\Sc_{\ell,k}|}}  \Fm_{\ell,k} \nuv_{\ell, k},  
\end{align*}
where the index set  $\Sc_{\ell,k} \subseteq \{0,\ldots, M-1\}$ includes all integers $m$ such that $2\pi m/M \in \Theta_{\ell,k}$
(where angles are taken modulo $2\pi$), 
$\Fm_{\ell,k}$ is the submatrix extracted from the $M \times M$ unitary DFT matrix $\Fm$ by taking the columns indexed by $\Sc_{\ell,k}$,  
and  $\nuv_{\ell,k}$ is an $|\Sc_{\ell,k}| \times 1$ i.i.d. Gaussian vector with components 
$\sim \Cc\Nc(0,1)$. Therefore,  $\hv_{\ell,k}$ is a Gaussian zero-mean random vector confined in the subspace spanned by the columns 
of $\Fm_{\ell,k}$.  

This channel model corresponds to the single ring of scatterers located around the UE \cite{adhikary2013joint}, with suitable quantization of the angle domain according to the $M$-array resolution limit. 
While the one-ring scattering model may be restrictive and is used in the simulations for simplicity, the fact that the dominant channel subspace is spanned by a collection of (possibly non-adjacent) columns of an $M\times M$ unitary DFT matrix is actually general enough for large uniform linear arrays (ULAs) and uniform planar arrays (UPAs). 
The angular support $\Sc_{\ell,k}$ imposes a simplified yet meaningful form of geometric consistency in the antenna correlation. Specifically, if two users are located within the same $\Delta$-wide angle concerning a RU, their channel vectors will share an identical subspace and therefore an identical covariance matrix. 
More precisely, when the spanned subspace has a dimension of 1 (i.e., generated by a single DFT column as in LOS situations), the channels of these two users will be co-linear with respect to the RU.

\subsection{Simulation Setup}

\begin{figure}
	\centerline{
		\includegraphics[width=0.4\linewidth, trim={19 34 18 19}, clip]{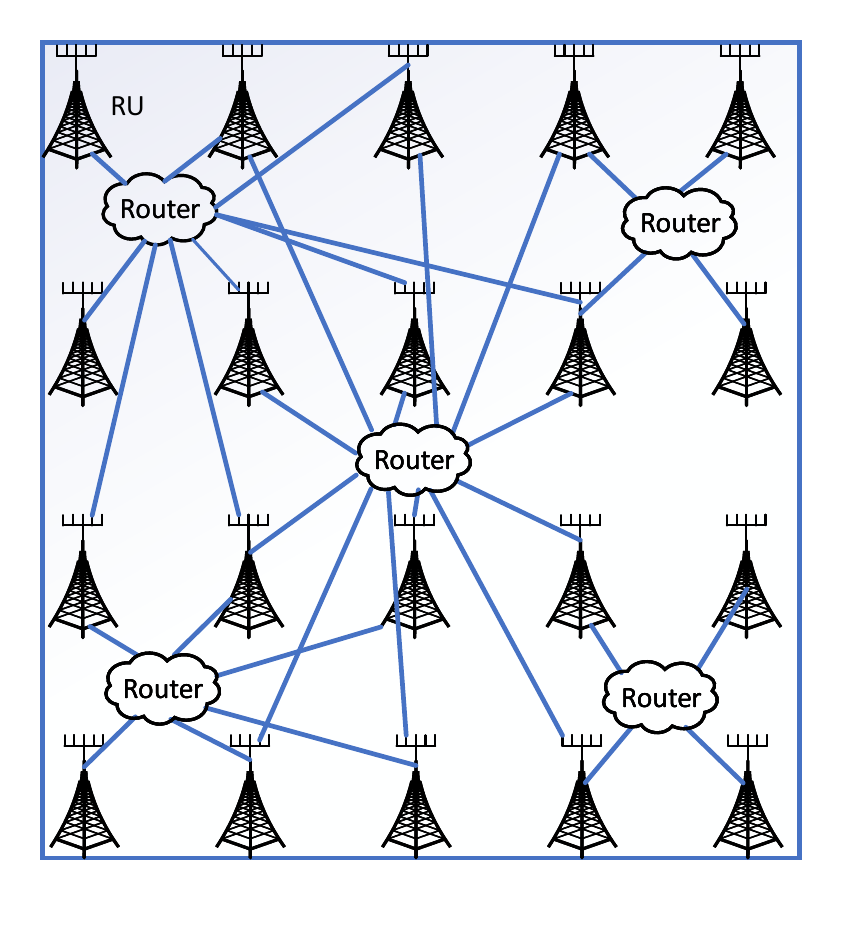} \hspace{.4cm}
		\includegraphics[width=0.4\linewidth, trim={19 34 18 19}, clip]{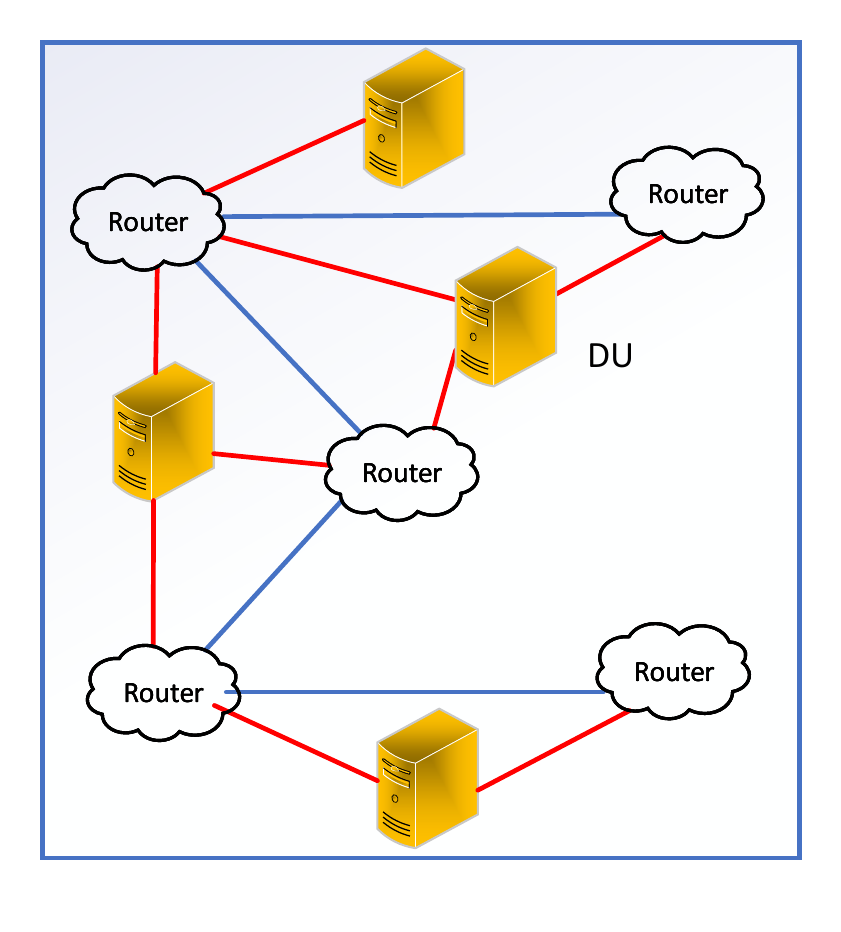}}
	\caption{ Left: The fronthaul links between RUs and routers. Right: The fronthaul links between routers and DUs.}
	\label{fig_fronthaul_connections}
\end{figure}

Consider a system with $L=20$ RUs and $K = 70$ UEs where each RU is equipped with $M = 10$ antennas and placed on a rectangular $5 \times 4$ grid and UEs are randomly and independently distributed in the area $A$.  
We set $\Delta = \pi/8$, $\taudmrs = 20$, the maximum cluster size $|\Cc_k| \leq \Cc_{\rm max} = 7$, 
and the SNR threshold $\eta = 1$ in (\ref{eq:snr_threshold}).
 We consider coherence blocks with
$T=200$ signal dimensions. The noise with power spectral density is $N_0 = -174 \text{ dBm/Hz}$, and the UL energy per symbol is $\bar{\beta} M \SNR = 1$ (i.e., 0 dB).
 The distance of expected pathloss $\bar{\beta}$ with respect to LOS and NLOS is $2.5 d_L$, where $d_L = \sqrt{\frac{A}{\pi L}}$ is the radius of a disk of area equal to $A/L$.
 The UL Tx power of all UEs depends on the RU density and the number of RU antennas. To achieve a certain level of coverage overlap among RUs, each UE is likely to associate with several RUs.
 The RUs are partially connected to $Q=5$ routers, which in turn are partially connected to $N=4$ DUs, as depicted in 
Fig.~\ref{fig_fronthaul_connections}. { For each of the DUs, we impose $Z_n = K/2 = 35$.}
 The optimization weights $\eta_L^{\rm ul}$, $\eta_Q^{\rm ul}$, $\eta_D^{\rm ul}$ for the UL, $\eta_L^{\rm dl}$, $\eta_Q^{\rm dl}$, $\eta_D^{\rm dl}$ for the DL, and $\eta_L$, $\eta_Q$, $\eta_D$ for the joint optimization problem are all set to $1$.  
In each coherence block,  $T - \taudmrs$ out of the $T$ signal dimensions are used for data transmission in the PHY layer. 
 The total UL/DL SE, i.e., the UL/DL PHY SE in bits per channel use (or bit/s/Hz) of all users, 
is thus given by 
\begin{align}
	{\rm SE}^{\rm ul} &= \left(1 - { {\gamma_{\rm DL}}} \right) \left( 1- \frac{\taudmrs}{T} \right) \sum_{k \in \Kc}  R^{\rm ul}_k \; \; \; \text{bit/s/Hz} ,  \label{eq:sum_phy_rate}
	\\
	{\rm SE}^{\rm dl} &= { {\gamma_{\rm DL}}}  \left( 1- \frac{\taudmrs}{T} \right) \sum_{k \in \Kc}   R^{\rm dl}_k \; \; \; \text{bit/s/Hz}, \label{eq:sum_phy_rate-dl}
\end{align}
where $R^{\rm ul}_k$ and $R^{\rm dl}_k$ are given by \eqref{ergodic_rate_ul} and  \eqref{ergodic_rate_dl}, respectively. 
The expectation in \eqref{ergodic_rate_ul} and \eqref{ergodic_rate_dl} is computed over $100$ channel realizations.

Note that for a given fronthaul quantization distortion $D$ in the UL, if the variance $\sigma_{\ell,k}^2$ of the observation for user $k$ at RU $\ell \in \Cc_k$ is less than $D$, RU $\ell$ sends zero bits to the cluster processor $\Cc_k$. This is equivalent to excluding RUs from the UL clusters. Recall that the DL information bits are encoded and precoded at the RUs. As a result, when $D/\sigma^2_{\min} > 1$, more and more RUs will be removed from the user-centric clusters, both in the UL and DL.

\subsection{System Evaluation for Different Optimization Approaches}

\begin{figure}[t!]
	\centering
	\begin{subfigure}{0.49\linewidth}
		\includegraphics[width=\linewidth]{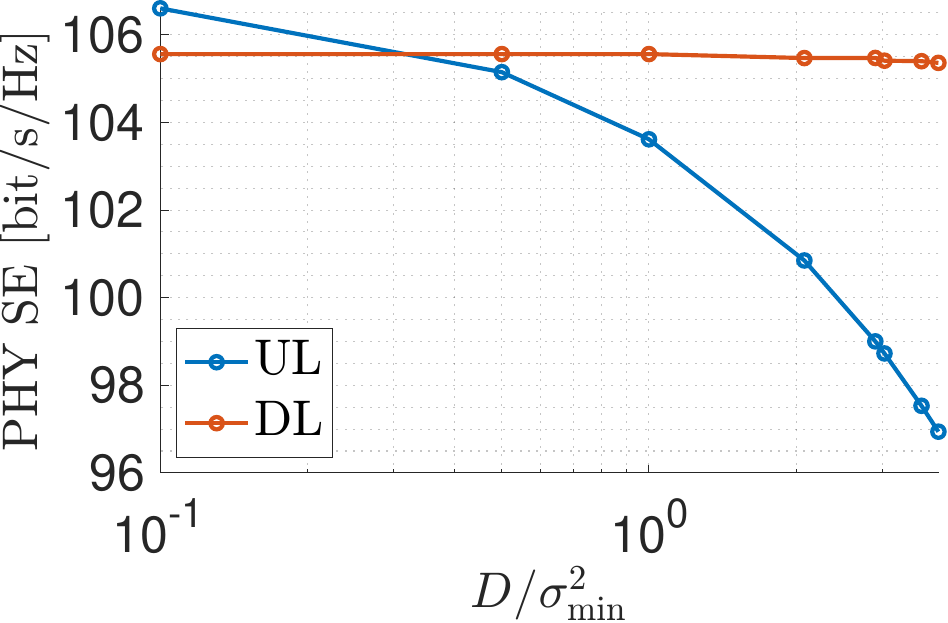}
		\caption{}
	\end{subfigure}
	\begin{subfigure}{0.49\linewidth}
		\includegraphics[width=\linewidth]{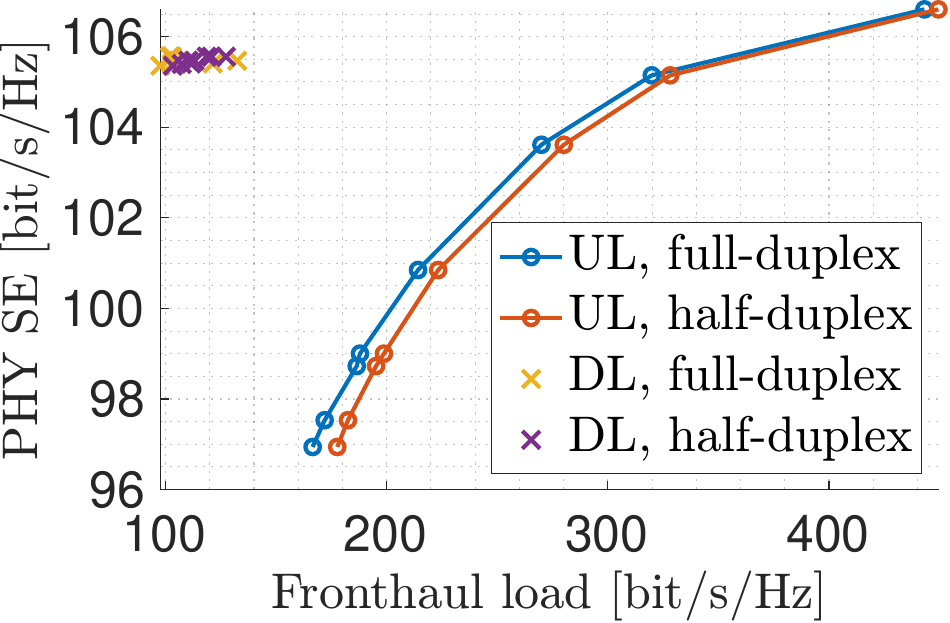}
		\caption{}
	\end{subfigure}
	\caption{ UL/DL PHY SE vs. distortion level (a) and corresponding fronthaul UL/DL load (b).}
	\label{PHY_rates_vs_D_and_fh_load}
\end{figure}

We start our numerical investigation of the UL/DL sum SE and fronthaul load under different distortion levels $D$ employing the  
proposed optimization problems for ${ {\gamma_{\rm DL}}} = 0.5$ (i.e., balanced UL-DL data traffic).
The results obtained by solving \eqref{opt_problem_fd} and \eqref{opt_problem_hd} are labeled as ``full-duplex'' and ``half-duplex'' in the figures, respectively. We define $\sigma^2_{\min} \defeq \min_{(\ell,k) \in \Ec_{\rm ran}}
\sigma_{\ell,k}^2$. Then,  all UE-RU associations in $\Ec_{\rm ran}$ are actually used for the UL signal processing when $D/\sigma^2_{\min} < 1$, while
as explained before, $D/\sigma^2_{\min}$ grows to values larger than 1, some RUs may be dropped from some clusters (since
their observations may be quantized with zero bits).  
 
 \begin{figure}[t!]
	\centering
	\begin{subfigure}{0.49\linewidth}
		\includegraphics[width=\linewidth]{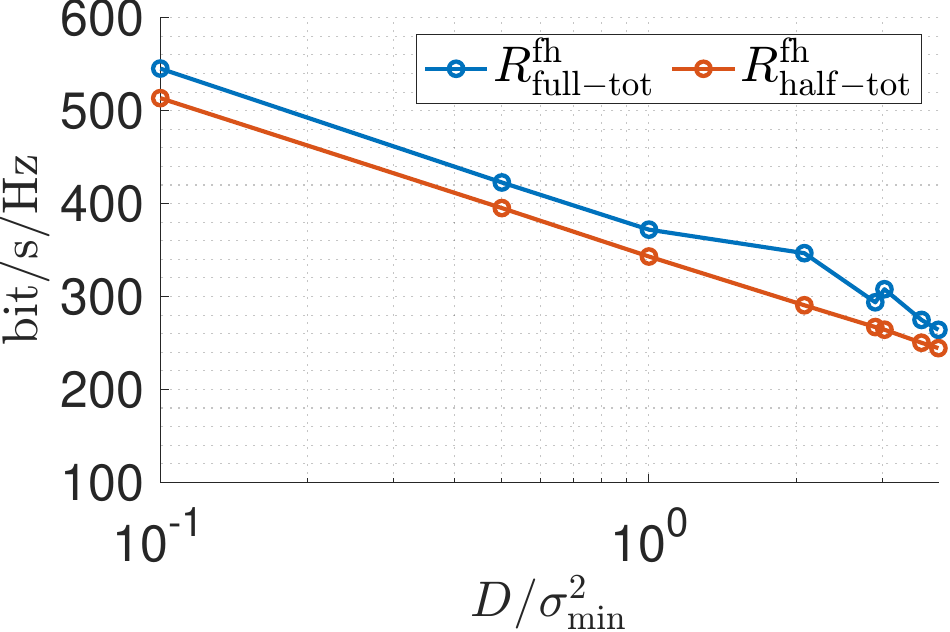}
		\caption{}
	\end{subfigure}
	\begin{subfigure}{0.49\linewidth}
		\includegraphics[width=\linewidth]{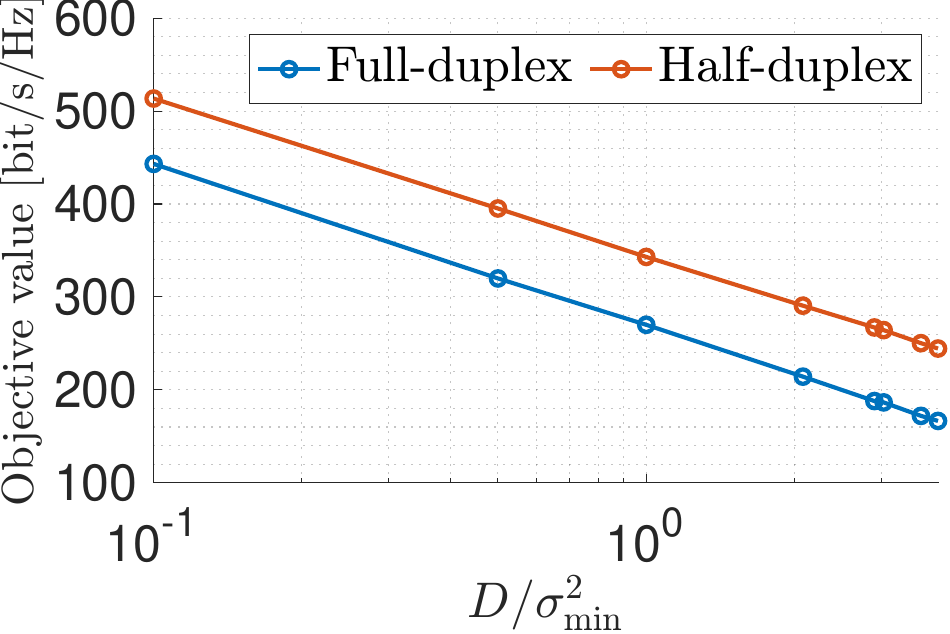}
		\caption{}
	\end{subfigure}
	\caption{ Total fronthaul load (a) and optimization objective function (b) vs. fronthaul quantization distortion level.}
	\label{PHY_rates_vs_OBJ_functions}
\end{figure}

Fig.~\ref{PHY_rates_vs_D_and_fh_load} illustrates the impact of $D$ on the UL/DL PHY SE and fronthaul load.  
The PHY SE is identical for both full-duplex and half-duplex fronthaul,  
since the optimization problems \eqref{opt_problem_fd} and \eqref{opt_problem_hd} only consider the fronthaul load.
We observe that increasing $D$ significantly affects both the UL PHY SE and the UL fronthaul load. 
In contrast, the impact of $D$ on the DL is essentially negligible, since the DL fronthaul carries the user information bits and the coding/modulation and precoding/combining is performed in the RUs.~\footnote{We notice here that in the O-RAN standard \cite[Section 4.2.1]{alliance2022ran} the precoding operations can be done either in the DUs or in the RUs, depending on the RU category. 
	The channel estimation and data detection in the UL are done at the DUs according to \cite[Section 4.2.3]{alliance2022ran}. Our results therefore call for a better design, where
	``smarter'' RUs 
	can locally encode and combine signals in both UL and DL transmissions. Notice that this is in line with the original idea of CF-mMIMO
	in \cite{Cell-Free}, based on local channel state information and local combining at the distributed antennas.}
  
For the full-duplex fronthaul case, we define the UL and DL fronthaul capacity as
\begin{align}
 R^{\rm fh}_{\rm full-UL} = (1 - { {\gamma_{\rm DL}}})  (\eta_L^{\rm ul} C_L^{\rm ul} + \eta_Q^{\rm ul} C_Q^{\rm ul} + \eta_D^{\rm ul} C_D^{\rm ul}), \label{eq_fd_ul_fronhaul_cap} 
\end{align}
and 
\begin{align}
R^{\rm fh}_{\rm full-DL} = { {\gamma_{\rm DL}}}  (\eta_L^{\rm ul} C_L^{\rm ul} + \eta_Q^{\rm ul} C_Q^{\rm ul} + \eta_D^{\rm ul} C_D^{\rm ul}), \label{eq_fd_dl_fronhaul_cap} 
\end{align}
respectively. 
In the half-duplex case, we define the  UL and DL fronthaul capacity as
 \begin{align}
 R^{\rm fh}_{\rm half-UL} =  \max_{\ell, q} \eta_L  \sum_{k}x_{k}^{\text{ru}}(\ell, q) &+ \max_{q, q'} \eta_Q \sum_{k}x_{k}^{\text{fh}}(q, q') \notag
 \\ &+ \max_{q, n} \eta_D \sum_{k}x_{k}^{\text{du}}(q, n), \label{eq_hd_ul_fronhaul_cap}
 \end{align}
 and
 \begin{align}
 R^{\rm fh}_{\rm half-DL} =  \max_{\ell, q} \eta_L  \sum_{k}y_{k}^{\text{ru}}(q, \ell) &+ \max_{q, q'} \eta_Q \sum_{k}y_{k}^{\text{fh}}(q, q') \notag
 \\ &+ \max_{q, n} \eta_D \sum_{k}y_{k}^{\text{du}}(n, q), \label{eq_hd_dl_fronhaul_cap}
 \end{align}
respectively.  Fig.~\ref{PHY_rates_vs_D_and_fh_load}(b)  shows UL/DL fronthaul capacities resulting from 
the proposed optimization as a function of $D$.
Furthermore, Fig.~\ref{PHY_rates_vs_D_and_fh_load}(b) shows that the UL is generally more demanding 
for the fronthaul load than the DL. 

For the UL, a $\approx 2.5$-fold increase of the fronthaul load from $\approx 170$ to $\approx 450$ bit/s/Hz corresponds to 
a rather small increase of $\approx 10$ \% of the PHY SE from $\approx 97$ to $\approx 106$ bit/s/Hz. 
This reveals that, from an engineering viewpoint, it is meaningful to employ a quite coarse resolution quantization (large $D$) in the 
UL fronthaul, since decreasing $D$ (i.e., increasing the fronthaul quantization rate) follows a law of diminishing returns. 

When considering the UL and DL fronthaul load separately as in Fig.~\ref{PHY_rates_vs_D_and_fh_load}, the full-duplex approach leads to a slightly smaller UL fronthaul load (and very similar DL fronthaul load). However, Fig.~\ref{PHY_rates_vs_OBJ_functions}(a) reveals that the total fronthaul 
load, given by 
 $R^{\rm fh}_{\rm full-tot} = R^{\rm fh}_{\rm full-UL}+ R^{\rm fh}_{\rm full-DL}$
  and 
by 
 $R^{\rm fh}_{\rm half-tot} = R^{\rm fh}_{\rm half-UL} \!+ R^{\rm fh}_{\rm half-DL} \!= \eta_L C_L + \eta_Q C_Q + \eta_D C_D$
in the full- and half-duplex cases, respectively, 
is smaller for the half-duplex case. This follows from the fact that the half-duplex optimization \eqref{opt_problem_hd}
minimizes directly the fronthaul sum capacity, while the full-duplex optimization \eqref{opt_problem_fd} minimizes the maximum of the 
UL and DL fronthaul capacities. 

Notice that this is not an {\em artifact} of our problem formulation, but in fact it is a fundamental fact: with full-duplex links, the UL and DL traffic are on orthogonal resources and one cannot trade one for the other, while with the half-duplex case a more flexible allocation is possible. 
This flexible allocation is automatically optimized by problem \eqref{opt_problem_hd}. 
Notice also  the non-monotonic behavior of the sum fronthaul load in the full-duplex case in  Fig.~\ref{PHY_rates_vs_OBJ_functions}. 
Since problem   \eqref{opt_problem_fd} minimizes the maximum of the UL and DL fronthaul capacities, the sum is not necessarily monotonic with respect to $D$ (in fact, the optimization objective function is actually monotonically decreasing, as seen in Fig.~\ref{PHY_rates_vs_OBJ_functions}(b)). 

\subsection{Optimal Distortion Level for UL-DL Traffic Imbalance}

\begin{figure}[t!]
	\centering
	\begin{subfigure}{0.49\linewidth}
		\includegraphics[width=\linewidth]{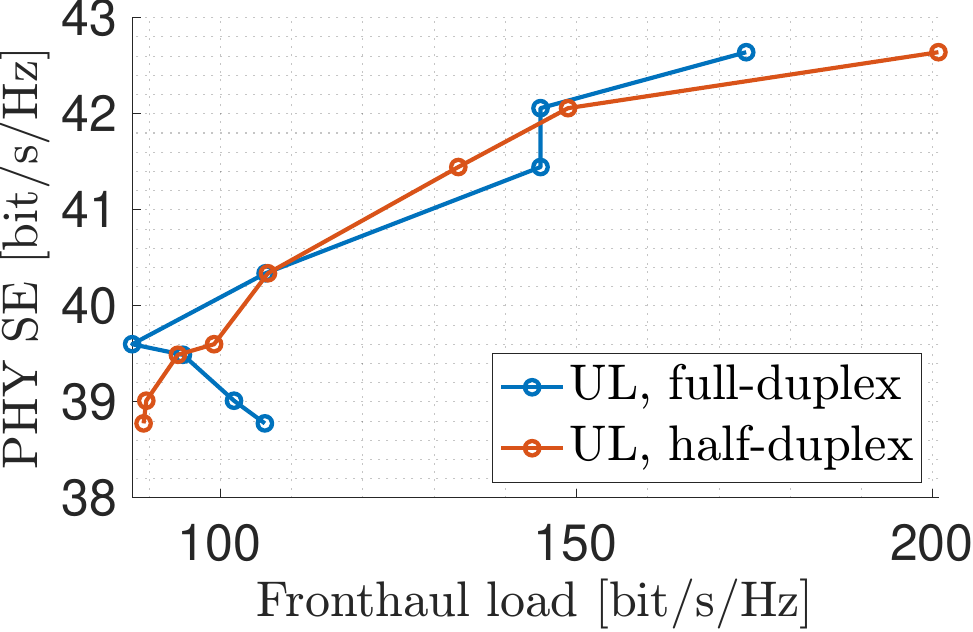}
		\caption{}
	\end{subfigure}
	\begin{subfigure}{0.49\linewidth}
		\includegraphics[width=\linewidth]{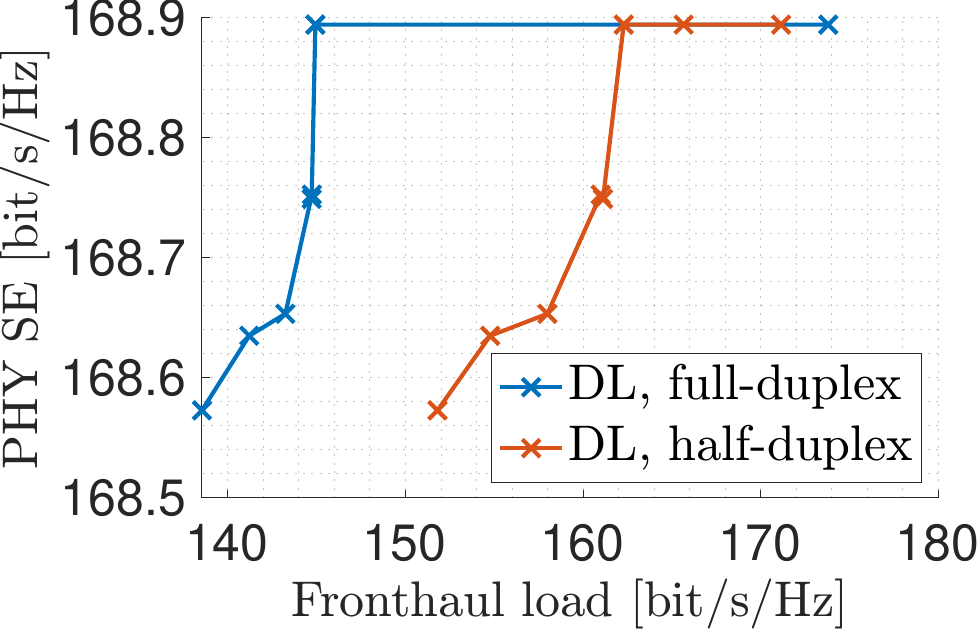}
		\caption{}
	\end{subfigure}
	\caption{UL/DL PHY SE vs. fronthaul UL/DL load. }
	\label{UL_DL_imb_PHY_vs_FH}
\end{figure}
 \begin{figure}[t!]
	\centering
	\begin{subfigure}{0.49\linewidth}
		\includegraphics[width=\linewidth]{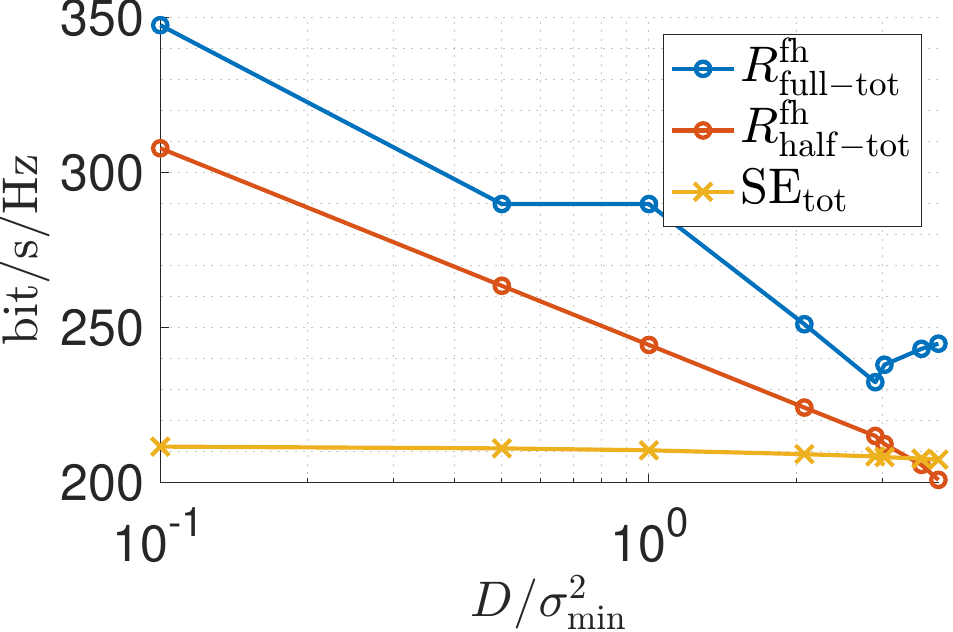}
		\caption{}
	\end{subfigure}
	\begin{subfigure}{0.49\linewidth}
		\includegraphics[width=\linewidth]{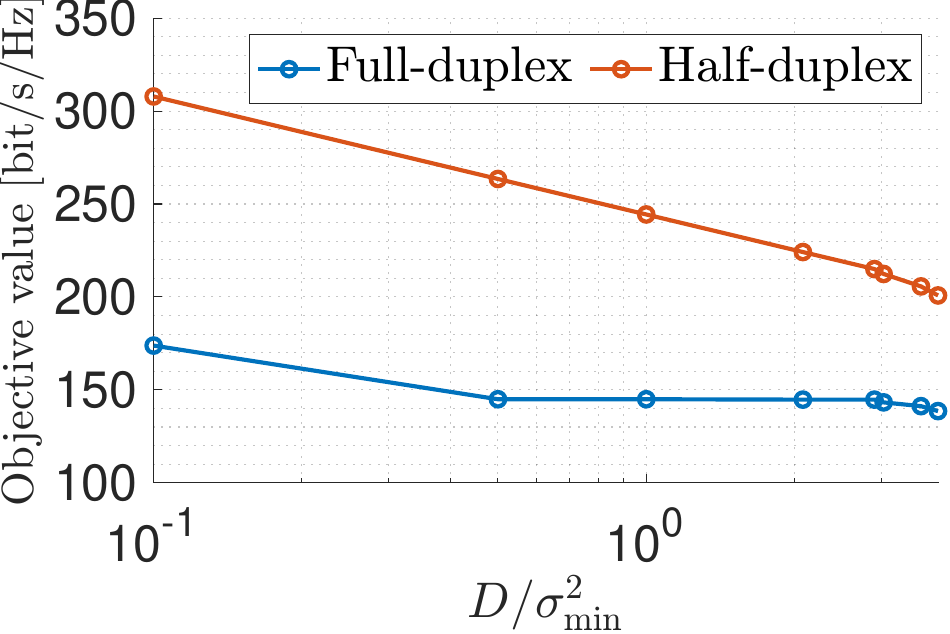}
		\caption{}
	\end{subfigure}
	\caption{Sum PHY SE, total fronthaul load (a) and optimization objective function (b) vs. distortion level. }
	\label{UL_DL_imb_FH_vs_D}
\end{figure}
In wireless networks, an imbalance between UL and DL traffic is common, with more data typically being sent in the DL than in the UL. 
Here we assume an UL-DL imbalance factor of ${ {\gamma_{\rm DL}}} = 0.8$.  
 Fig.~\ref{UL_DL_imb_PHY_vs_FH} shows the PHY SE versus fronthaul capacity for the UL and DL, respectively, for both full and half-duplex fronthaul 
cases.  
Notice that in this case the DL SE is much larger, due to the large fraction of timeslots dedicated to the DL. 
Also, both UL and DL SE are very insensitive to the fronthaul capacities and have an essentially flat behavior in the considered range of
fronthaul capacity. 
Since the DL fronthaul is dominant for small $D$ (i.e., low PHY SE), the UL fronthaul load is thus not necessarily monotonic with respect to the PHY SE in this regime with full-duplex links (see Fig.~\ref{UL_DL_imb_PHY_vs_FH}(a)).

Fig.~\ref{UL_DL_imb_FH_vs_D}(a) shows the full-duplex and half-duplex total fronthaul load and the sum 
PHY SE, defined as $ {\rm SE}_{\rm tot} = {\rm SE}^{\rm ul} + {\rm SE}^{\rm dl}$. Since 
${\rm SE}_{\rm tot}$ is almost constant with $D$, while $R^{\rm fh}_{\rm full-tot}$ and $R^{\rm fh}_{\rm half-tot}$ decrease as $D$ increases (except for the slight increase at large $D$ in the full-duplex case), 
it is clear that a coarse fronthaul quantization is a good design choice. Of course, if $D$ increases further, eventually the
UL PHY SE will drastically decrease and eventually drop to zero, but this is not a relevant regime for the system and
therefore it is not taken into consideration. 
The observed rise of $R^{\rm fh}_{\rm full-tot}$ for large $D$ can be attributed to the dominance of either UL or DL fronthaul traffic, which defines the overall objective value. The load in the non-dominant fronthaul direction only needs to be below that of the dominant fronthaul direction. 
With $\gamma_{DL} = 0.8$ in the wireless segment, 
the DL fronthaul load dominates for all values of $D$ (see Fig.~\ref{UL_DL_imb_PHY_vs_FH}). Initially, the UL fronthaul load experiences a decrease with larger $D$, but eventually exhibits a small increase. Importantly, this increase does not impact the objective value of the full-duplex problem as evidenced in Fig.~\ref{UL_DL_imb_FH_vs_D}(b). 
Since the DL fronthaul load decreases only slightly  for larger $D$, the sum UL/DL fronthaul load experiences an increase. 
However, it is important to note that the vertical coordinate in Fig.~\ref{UL_DL_imb_FH_vs_D}(a) does not directly correspond to the objective function of the full-duplex case.

\section{Conclusion}
\label{sec:conclusion} 

In this work, we proposed a novel four-tier scalable cell-free user-centric wireless network model that takes explicitly into account
the fronthaul topology, the computation constraint of DUs, and the quantization of the received signal at the RUs. 
The model is closely inspired by the O-RAN framework, with the difference that we consider ``smarter'' RUs that can locally combine and precode signals in the UL and DL. 
This aspect of the considered system does not follow the current O-RAN paradigm, but should be considered as 
a smart alternative and more advanced scheme.
Each user is associated with a finite size user-centric cluster of RUs, and in the UL, we employ local LMMSE detectors and cluster-level combining under fronthaul quantization. 

We formulated the joint optimization of the fronthaul load (routing flow problem) and of the placement of the user-centric cluster processors in the DUs as mixed-integer linear programs, that can be efficiently and exactly solved with standard tools even for relatively large systems.  
We consider two scenarios: 1) min-max UL and DL optimization is performed for full duplex fronthaul links; 2) 
UL and DL are jointly optimized under the assumption of half-duplex fronthaul. 

We have carefully characterized the effect of quantization distortion in the UL through an information-theoretic rate-distortion argument, 
which is closely achievable in practice using entropy coded scalar dithered quantization. This is relevant for the UL, where 
quantized received signals must be sent from the RUs to the DU via the fronthaul. In contrast, no quantization is needed for the DL since
in this case the DUs can send directly the information bits, which are then encoded/modulated and combined with precoding coefficients in the RUs.

Our results show that for balanced TDD traffic the fronthaul bottleneck is represented by the UL, and it is convenient to use
a rather coarse fronthaul quantization since the increase of the quantization distortion 
yields a relatively small degradation of the UL PHY SE at the benefit of a greatly reduced UL fronthaul load. 
This effect is less and less evident for more and more imbalanced TDD traffic: when the DL consumes a much larger fraction of the transmission 
resources than the UL, the UL fronthaul load is not the major system constraint in any case and carefully optimizing the quantization distortion becomes less relevant. 

Beyond the specific results presented here, the merit of this work consists of the modeling and the formulation of the 
joint optimization framework for fronthaul load and distributed computation allocation, and the clean analytical framework
to study the effect of ``practical'' fronthaul quantization in the UL. 


\appendices

\section{Proof of Lemma \ref{lemmaQ} and Lemma \ref{lemmaR}} \label{AppendixQ}

We prove Lemma \ref{lemmaQ} in a few steps. First, we consider
the lossy source coding problem for i.i.d. discrete sources as in \cite{Cover:2006}. 
Let  $\{X_i\}$  denote a discrete memoryless source whose symbols take on values in a set $\Xc$ with (first-order) distribution $P_X$. Let $\widehat{\Xc}$ be a representation alphabet and consider a bounded distortion function $d : \Xc \times \widehat{\Xc} \rightarrow \RR_+$ such that, for any $(x,\widehat{x}) \in \Xc \times \widehat{\Xc}$, 
$d(x,\widehat{x}) \leq d_{\max} < \infty$.
The per-symbol distortion between 
two sequences $\xv \in \Xc^n$ and $\widehat{\xv} \in \widehat{\Xc}^n$ is defined as 
\begin{align*}
d(\xv, \widehat{\xv}) \defeq \frac{1}{n} \sum_{i=1}^n d(x_i, \widehat{x}_i).
\end{align*}
 Let $X \sim P_X$. 
Consider a representation random variable $\widehat{X} \in \widehat{\Xc}$ jointly distributed with $X$ as $(X, \widehat{X}) \sim P_X P_{\widehat{X}|X}$ for some conditional distribution $P_{\widehat{X}|X}$.
From the achievability proof of \cite[Th. 10.2.1]{Cover:2006}, we know that there exist lossy source coding schemes 
$(\phi_n,\psi_n)$ for the above problem achieving 
\begin{equation} 
	R_X(D) = \min_{P_{\widehat{X}|X} : \EEE[d(X,\widehat{X})] \leq D} \; I(X;\widehat{X}). \label{RDfunct-DMS}
\end{equation}
For some $\epsilon > 0$ and sufficiently large $n$, let $\Tc^{(n)}_\epsilon(X,\widehat{X})$ denote the 
strongly jointly typical set (see definition in \cite{elgamal_kim_2011}).~\footnote{In short, $\Tc^{(n)}_\epsilon(X,\widehat{X})$ is the set of all pairs of length-$n$ sequences of  $(\xv, \widehat{\xv}) \in \Xc^n \times \widehat{\Xc}^n$ such that their joint empirical distribution is close (within a relative error $\epsilon$) to $P_{X,\widehat{X}}$.}
From the proof of achievability of (\ref{RDfunct-DMS}), we know that 
for some $\epsilon > 0$ and given $n$,  if $R > I(X; \widehat{X}) + 3 \epsilon$ 
the distortion $D_n = \EE[ d(X, \widehat{X})] + \epsilon + P_e^{(n)} d_{\max}$ is achieved 
with overwhelming high probability (as $n$ increases) by the following random coding scheme:
\begin{itemize}
	\item {\bf Random codebook generation:} generate codewords $\widehat{X}^n(m)$ for $m = 1, 2, \ldots, 2^{nR}$ 
	of length-$n$ with i.i.d. components $\sim P_{\widehat{X}}$; 
	\item {\bf Joint typicality encoder:} Define the index set $\Mc(X^n) \defeq \{ m : (X^n, \widehat{X}^n(m)) \in \Tc_\epsilon^{(n)}(X, \widehat{X}) \}$ containing the indices of the codewords jointly typical with the given source sequence $X^n$. If
	$\Mc(X^n) \neq \emptyset$, then $\phi_n(X^n)$ is any (arbitrarily chosen) index in $\Mc(X^n)$. Otherwise, 
	$\phi_n(X^n) = 1$. 
\end{itemize}
In particular, the probability of encoding error 
$P_e^{(n)} \leq \PP \left ( \Mc(X^n) = \emptyset  \right )$ 
vanishes doubly exponentially in $n$.  
 Letting $\epsilon \downarrow 0$ and $n \rightarrow \infty$, we have that any rate $R > I(X; \widehat{X})$ 
is achievable with distortion $D = \EE[ d(X, \widehat{X})]$. 

Next, we notice that 
\cite[Th. 10.2.1]{Cover:2006} generalizes to continuous-valued discrete memoryless sources and sufficiently well-behaved
distortion measures (e.g., real and complex valued finite variance sources and MSE distortion) in a completely standard way
(for details, see \cite{Cover:2006} and also \cite[Ch. 10]{elgamal_kim_2011}). 
This yields that we can generate a lossy source coding scheme by: a) choosing a ``template'' joint distribution 
$P_{X, \widehat{X}}$ for $(X, \widehat{X})$ such that the $X$-marginal coincides with 
the original distribution of the source $P_X$; b) using joint typicality encoding. 
As a result, the achieved distortion will be as close as desired to $D = \EE[ | X - \widehat{X}|^2]$ provided that the rate $R$ is not smaller than $I(X; \widehat{X})$. 

Finally, notice that for a stationary ergodic source $\{X_i\}$, since the marginal distribution $P_X$ of its elements $X_i$ does not depend on $i$, the very same achievability result still holds.~\footnote{Notice that in this case
	the true rate-distortion function of the source may be smaller than (\ref{RDfunct-DMS}), since the latter somehow ``ignores'' the time-dependency of the source. Nevertheless, the rate in (\ref{RDfunct-DMS}) is achievable by the very same approach
	outlined for memoryless sources.}

Now, if $D \geq \sigma^2$, then by encoding any source sequence into the all-zero sequence (requiring zero bits)
yields distortion $\frac{1}{n} \EE[\|X\|^2] = \sigma^2 \leq D$. In this trivial case, the rate to achieve distortion (not larger than) $D$ is zero. Considering the non-trivial case $D < \sigma^2$,
by choosing $P_{X,\widehat{X}}$ induced by the model $X = \widehat{X} + Q_D$, where $Q_D \sim \Cc\Nc(0, D)$ is independent of $\widehat{X}$, the MSE distortion $\EE[|X - \widehat{X}|^2] = \EE[|Q_D|^2] = D$ can be achieved at any rate
$R > I(X; \widehat{X})$. In particular, we have
\begin{eqnarray} 
	I(X; \widehat{X})  & = & h(X) - h(X | \widehat{X}) \nonumber \\
	& = & h(X) - h(X - \widehat{X} | \widehat{X}) \nonumber \\
	& = & h(X) - h(Q_D | \widehat{X}) \nonumber \\
	& = & h(X) - h(Q_D) \nonumber \\
	& \leq & \log_2( \pi e \sigma^2) - \log_2 (\pi e D) = \log_2 ( \frac{\sigma^2}{D} ). \; \; \; \; \;
\end{eqnarray}
Hence, Lemma \ref{lemmaQ} is proved. 

In order to prove Lemma \ref{lemmaR}, we use the following fact. Consider the coding schemes in Lemma \ref{lemmaQ}, with random codebook generation according to the template joint probability model $P_{X,\widehat{X}}$ induced by
$X = \widehat{X} + Q_D$. Let $\widehat{m} = \phi_n(X^n)$ and $\widehat{X}^n = \widehat{X}^n(\widehat{m})$, such that 
$(X^n, \widehat{X}^n)$ are the random source sequence and its corresponding quantization codeword, under
the joint typicality encoder $\phi_n$ defined above. 
By the joint typicality condition, we have that, with overwhelming high probability with respect to the random codebook generation, the empirical joint distribution of the $i$-th symbol pair $(X_i, \widehat{X}_i)$, for any $i$,
is arbitrarily close to the chosen template joint distribution $P_{X,\widehat{X}}$.
In particular, this implies that $X_i = \widehat{X}_i + Q_{D, i}$, where $Q_{D, i} \sim \Cc\Nc(0,D)$ and 
$\widehat{X}_i$ and $Q_{D, i}$ are  statistically independent. 

Next, we consider the linear MMSE estimation of $\widehat{X}_i$ from $X_i$ given by 
\[ \widehat{X}_i^{\rm mmse} = \frac{\EE[ \widehat{X}_i X_i^*]}{\EE[|X_i|^2]} X_i = 
\frac{\EE[|\widehat{X}_i|^2]}{\EE[|X_i|^2]} X_i = \alpha X_i, \]
where, by noticing that 
\[ \sigma^2 = \EE[|X_i|^2] = \EE[|\widehat{X}_i + Q_{D, i}|^2] = \EE[|\widehat{X}_i|^2] + D, \]
$\alpha$ is immediately seen to be given by (\ref{bussgangalpha}). 
Define the resulting error variable as $E_i \defeq \widehat{X}_i - \alpha X_i$. By the well-known properties of the LMMSE estimator,  the error $E_i$ is uncorrelated with the ``observation'' $X_i$, and its variance 
(i.e., the resulting estimation MSE) is given by 
\begin{eqnarray}
	\EE[|E_i|^2] & = & \EE[|\widehat{X}_i - \alpha X_i|^2] \nonumber \\
	& = & \EE[\widehat{X}_i (\widehat{X}_i - \alpha X_i)^*] \nonumber \\
	& = & \EE[|\widehat{X}_i|^2] - \alpha \EE[ \widehat{X}_i X_i^*] \nonumber \\
	& = & \sigma^2 - D - \alpha^2 \sigma^2  \nonumber \\
	& = & (1 - D/\sigma^2) D.
\end{eqnarray}
Hence, Lemma \ref{lemmaR} is proved. 

\bibliography{refs-CF-UC}

\end{document}